\documentclass[aps,prd,twocolumn,10pt,superscriptaddress,nofootinbib,nobibnotes,longbibliography]{revtex4-1}

\usepackage{amssymb}
\usepackage{graphicx}
\usepackage{amsmath}
\usepackage{hyperref}
\usepackage{subfigure}
\usepackage{multirow}
\usepackage{xcolor}
\usepackage{ulem}
\usepackage{setspace}

\newcommand{\md}{\mathrm{d}}
\newcommand{\me}{\mathrm{e}}
\newcommand{\pll}{\parallel}

\newcommand{\Rmnum}[1]{\expandafter\@slowromancap\romannumeral #1@}

\begin{document}

\title{\boldmath Squared-field cross-correlation between kinetic Sunyaev-Zel'dovich effect and 21-cm intensity mapping}

\author{Zi-Yan Yuwen}
\affiliation{Department of Physics, Stellenbosch University, Matieland 7602, South Africa}
\affiliation{Institute of Theoretical Physics, Chinese Academy of Sciences (CAS), Beijing 100190, China}
\affiliation{University of Chinese Academy of Sciences (UCAS), Beijing 100049, China}

\author{Yu-Er Jiang}
\affiliation{Department of Physics, Stellenbosch University, Matieland 7602, South Africa}
\affiliation{University of Chinese Academy of Sciences (UCAS), Beijing 100049, China}
\affiliation{National Astronomical Observatories, Chinese Academy of Sciences, Beijing 100012, China}

\author{Yin-Zhe Ma}
\email{Corresponding author: mayinzhe@sun.ac.za}
\affiliation{Department of Physics, Stellenbosch University, Matieland 7602, South Africa}

\author{Paul La Plante}
\affiliation{Department of Computer Science, University of Nevada, Las Vegas, Nevada 89154, USA}
\affiliation{Nevada Center for Astrophysics, University of Nevada, Las Vegas, Nevada 89154, USA}

\author{Adam Lidz}
\affiliation{Center for Particle Cosmology, Department of Physics and Astronomy, University of Pennsylvania, Philadelphia, Pennsylvania 19104, USA}

\author{Yan Gong}
\affiliation{University of Chinese Academy of Sciences (UCAS), Beijing 100049, China}
\affiliation{National Astronomical Observatories, Chinese Academy of Sciences, Beijing 100012, China}
\affiliation{Science Center for China Space Station Telescope, National Astronomical Observatories, Chinese Academy of Sciences, Beijing 100012, China}

\begin{abstract}
Neutral hydrogen (HI) 21-cm intensity mapping is an effective method to track the distribution of baryonic matter, and extract astrophysical and cosmological information. The 21-cm intensity field has a nonvanishing cross-correlation with the kinetic Sunyaev-Zel'dovich (kSZ) effect that traces the velocity and density perturbations of free electrons. By using the linear perturbation theory, in this paper we calculate analytically, for the first time, the cross-correlation between the squared kSZ field and the projection of the squared HI intensity mapping field with the flat-sky approximation. This statistic remains nonvanishing even after the long-wavelength line-of-sight modes ($k_{\parallel}$) are removed due to foreground contamination. We further forecast for the prospects of detection with the SKA-MID 21-cm intensity mapping experiments (redshifts in range of $0.3 < z < 1$), and the kSZ maps measured by the Atacama Cosmology Telescope (ACT) and Simons Observatory (SO). The predicted cumulative signal-to-noise ratio is $1.92$ for SKA-ACT and $3.99$ for SKA-SO. These results show a possible on-the-edge detection on the cross-correlation signal at low redshifts, which in turn could serve as a validation step toward using it for the Epoch of Reionization studies.
\end{abstract}
\maketitle

\section{Introduction}
\label{sec:introduction}
The standard model of cosmology $\Lambda$CDM~\cite{Efstathiou:1990xe} has been well established by various astronomical observations from the Type-Ia supernovae~\cite{Riess1998,Perlmutter1998}, cosmic microwave background (CMB)~\cite{Planck:2018vyg,ACT:2020frw,Camphuis2025} and the large-scale structure (LSS)~\cite{DESI:2016fyo,Euclid:2024yrr} of the universe in the past two decades. Apart from the successful space-borne CMB observation project \textit{Planck}~\cite{Planck:2018vyg}, there are many ongoing ground-based CMB observations such as Atacama Cosmology Telescope (ACT;~\cite{ACT:2020frw,ACTPol:2015teu}), South-Pole Telescope (SPT;~\cite{Camphuis2025}), and Simons Observatory (SO; ~\cite{SimonsObservatory:2025wwn}), CMB-High Definition (CMB-HD~\cite{Sehgal:2019ewc}) etc., with much higher angular resolution and lower pixel noises that enable the small-scale CMB measurements in better accuracy. In recent years, {\it Planck}, ACT, and SPT have jointly measured the temperature anisotropy power spectrum out to $\ell \simeq 4000$, which enters into the regime where the secondary CMB anisotropies become dominant. Among these effects, the kinematic Sunyaev-Zel'dovich effect (kSZ~\cite{Sunyaev1980}) serves as an important tracer for the distribution of baryonic matter. The kSZ effect arises from inverse Compton scattering of CMB photons by the free electrons with peculiar velocities, which can be written as
\begin{eqnarray} 
\label{eq: deltaT kSZ / TCMB}
\frac{\delta T_{\mathrm{kSZ}}(\hat{n})}{T_{\mathrm{CMB}}} &=& -\int \md \chi~ g(\chi) \hat{n}\cdot \vec{q}(\chi\hat{n}),
\end{eqnarray}
where $\vec{q}=(1+\delta_\mathrm{e})\vec{v}$ is the peculiar momentum, $\delta_\mathrm{e}$ is the electron density contrast, $\chi(z)$ is the comoving distance to redshift $z$, and the peculiar velocity $\vec{v}$ has been normalized with the speed of light. The visibility function $g(\chi)$ can be written as a function of redshift
\begin{align} \label{eq: g(z)}
    g(\chi) &= \me^{-\tau(z)} \frac{\md \tau}{\md \chi} \\ &= \sigma_\mathrm{T} \frac{\bar{n}_{\mathrm{e}}(z)}{1+z} \exp\left(-\sigma_\mathrm{T}\int_0^z \frac{\bar{n}_{\rm e}(z')}{1+z'} \frac{c ~\md z'}{H(z')}\right), 
\end{align}
where $\tau$ is the optical depth, and $\sigma_\mathrm{T}$ is the Thomson cross-section. $\bar{n}_{\rm e}(z)$ is the mean electron density at redshift $z$, which is equal to
\begin{eqnarray}
\bar{n}_{\rm e}(z) &=& \bar{n}_{\mathrm{b},0} X_\mathrm{e}(z) (1+z)^3 \nonumber \\
&=& \left(\frac{\rho_{\rm cr}\Omega_{\rm b}}{\mu_{\rm e}m_{\rm p}}\right)X_\mathrm{e}(z) (1+z)^3,
\end{eqnarray}
where $\bar{n}_{\mathrm{b},0}$ is the present day's mean number density of baryons, $X_\mathrm{e}(z)$ is the mean ionized fraction at redshift $z$, $\rho_{\rm cr}=1.879h^{2}\times 10^{-29}\,{\rm g}\,{\rm cm}^{-3}$ is the critical density of the Universe, $\Omega_{\rm b}$ is the fractional baryon density, and $\mu_{\rm e}\simeq 1.14$ is the mean mass per electron~\cite{Ma2014}.

The kSZ effect is often used in conjunction with galaxy cluster data to study the missing baryons problem~\cite{Planck2016-unbound,Lim:2017,McCarthy:2024nik} and reconstruct velocity fields~\cite{Ma2017,Ma2018,Li2018,ACTPol:2015teu,DES:2023mug,Zhou:2025fgv}. Various techniques have been investigated to detect the cross-correlation between the kSZ effect and other large-scale structure tracers, such as the pairwise momentum field~\cite{Hand2012,Planck2016-unbound,Li2018}, projected density field~\cite{Hill:2016dta,Ferraro:2016ymw}, galaxy clusters~\cite{Planck2018-dispersion}, and the weighted reconstructed velocities~\cite{Hadzhiyska:2024qsl}.

In recent years, researchers have considered potential cross-correlations between the patchy reionization-era kSZ signal and other tracers of the Epoch of Reionization (EoR)
~\cite{Hotinli:2020csk,Begin:2021skl,Georgiev:2023yqr,Zhou:2025fgv}. Among all the large surveys, neutral hydrogen (HI) 21-cm intensity mapping (IM) survey becomes one of the most powerful large-scale structure surveys that can cover wide redshift ranges and large sky areas with efficient scanning strategies~\cite{Lidz:2011dx,Battye:2012tg,Jiang:2023zex}. In recent years, Green Bank Telescope~\cite{Chang2010,Masui2013}, Parkes Telescope~\cite{Pen2009,Anderson2018,Tramonte2019,Tramonte2020}, and MeerKAT telescope~\cite{Cunnington2022} have measured the HI IM to a higher and higher precision, and it is almost certain that the future Square Kilometre Array (SKA) will provide a large survey sample of the HI intensity maps which will constrain cosmological parameters with high precision~\cite{MeerKLASS:2017vgf,SKA:2018ckk}.


In this paper, we consider the cross-correlation between the squared kSZ field and the squared HI intensity field.
In each case, it is necessary to square the fields before cross-correlating them. 
For the kSZ effect, because of the statistical isotropy, clusters with positive and negative peculiar velocities have equal weights. Therefore, the direct two-point correlation function $\langle \delta T_{\mathrm{kSZ}}(\vec{k}_1) \delta T_{\mathrm{HI}}(\vec{k}_2) \rangle$ simply vanishes given enough statistics, which can be avoided by taking the square of the kSZ field~\cite{Dore:2003ex}. In practice, a Wiener filter will be applied before squaring the field to suppress the primary CMB, lensing effects, residual foreground contamination, and detector noise.
In the case of the 21-cm field, without squaring, the relevant signal would involve the bispectrum $\langle \delta T_{\mathrm{kSZ}}(\vec{k}_1)\delta T_{\mathrm{kSZ}}(\vec{k}_2) \delta T_{\mathrm{HI}}(\vec{k}_3) \rangle$, which is proportional to $\delta_{\mathrm{D}}^{(3)}(\vec{k}_1 + \vec{k}_2 + \vec{k}_3)$. Since the kSZ effect is a projected signal, this statistic only receives contributions with $k^\parallel_1$ and $k^\parallel_2$ nearly equal to zero, while the Dirac Delta function enforces $k^{\parallel}_{3} = 0$, as required by the translational invariance. 
This means that only the low $k^{\pll}$ modes in the HI field can contribute to the estimator. However, because the large-scale radial mode of the HI IM is heavily contaminated by the spectrally smooth foregrounds, the HI IM data after foreground removal procedure (either through PCA~\cite{Davis:1985rj}, FASTICA~\cite{Wolz:2013wna}, or GNILC~\cite{Dai2025} methods) tend to lose all low-$k^{\parallel}$ modes. Therefore, we need to square the HI IM cube so that we can fully utilize the high-$k^{\parallel}$ modes that are {\it not} removed by foreground removal procedure. In Appendix~\ref{app: PS v.s. SP}, we prove that, by taking the square of the HI IM field before projecting it along the line-of-sight we take advantage of all high-$k^{\parallel}$ modes that contribute to the cross-correlation signal. Therefore, our goal is to model cross-correlations between the kSZ-squared field and the projection of the 21-cm-squared field. This is a two-dimensional (2D) cross-correlation which can be performed at the map level.

On large scales, the kSZ effect carries imprints of ionized baryons and follows the distribution of dark matter, which is also the case for the HI field. Therefore, we expect to see a positive correlation function between the two on large scales. 
On small, galactic scales neutral hydrogen resides predominantly in dense, neutral regions which remain self-shielded against photoionizing radiation~\cite{Rahmati:2012rg,Liu:2022iyy}. In contrast, free electrons occupy the ionized phase of the circumgalactic medium (CGM), leading to an anticorrelation between the two tracers, which could be seen if the maps have sufficient resolution. However, this small-scale anticorrelation signal may be challenging to see because of the coarse resolution of foreseeable 21-cm measurements. Therefore, we focus our effort in predicting the signal on large scales with $\ell \lesssim 900$, and forecast its detectability.

The paper is organized as follows. In Sec.~\ref{sec:cross-correlation}, we provide a detailed derivation of the analytical result consisting of the projection method~\ref{subsec: projection}, general formulation of cross-correlations~\ref{subsec: cross-correlation calculation}, and the triple-spectrum evaluation~\ref{subsec: triple-spectrum}. In Sec.~\ref{sec:num_res} we show our numerical results of the cross-correlation~\ref{subsec: num cross} and the signal-to-noise ratio forecasts~\ref{subsec: SNR} by adopting instrumental parameters of SKA-MID and ACT/SO kSZ measurement. Section~\ref{sec:con} is devoted to conclusions and discussions.

Apart from the forecasting parameters, we adopt the best-fitting {\it Planck} 2018 ``TT+TE+EE+lowE+lensing+BAO'' cosmological parameters~\cite{Planck:2018vyg}  to compute the matter power spectrum, particularly, using $H_0=67.7\,{\rm km}\,{\rm s}^{-1}\,{\rm Mpc}^{-1}$, $\Omega_\mathrm{b}h^2 = 0.022$,  $\Omega_\mathrm{c}h^2 = 0.119$, $\tau=0.056$, $\ln(10^{10}A_{\rm s})=3.05$ and $n_{\rm s}$=0.966.

\section{Cross-correlation of the projection fields}\label{sec:cross-correlation}

In this section, we derive an analytical equation for the cross-correlation between the kSZ square field and the HI intensity mapping square field. We first go through the 2D projection procedure of a three-dimensional (3D) field under the flat-sky approximation, and then provide a detailed calculation. 

\subsection{Two dimension projection field}\label{subsec: projection}

In general, any 3D scalar field $\delta_X(\vec{x})$ can be projected to a 2D field $X(\hat{n})$ as a function of line-of-sight direction $\hat{n}$
\begin{align}\label{eq:projection2D}
    X(\hat{n}) =  B_X * \int \md \chi ~W_X(\chi) \delta_X(\chi,\chi \hat{n}),
\end{align}
where $\chi$ is the comoving distance to the observer, $W_X$ is the integration kernel for the $X$ field satisfying the normalization condition $\int \md \chi W_X(\chi) = 1$, and $B_X$ is the telescope beam function in the real space which is convolved with the projected field \cite{Dore:2003ex}.

\begin{figure}
    \centering
    \includegraphics[width=0.7\linewidth]{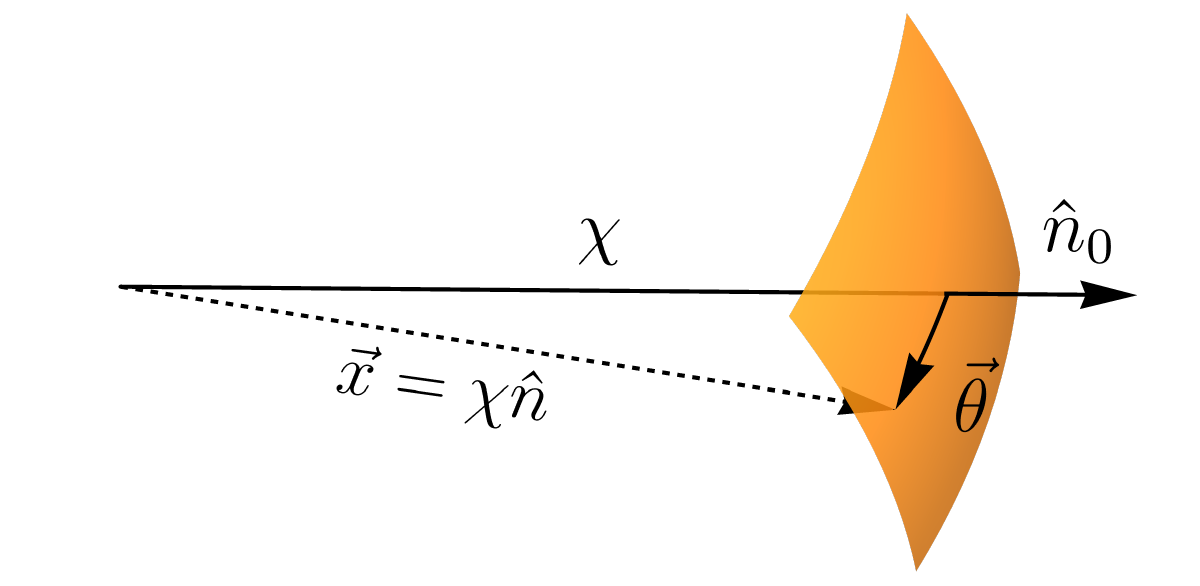}
    \caption{An intuitive picture and notation explanation for the flat-sky approximation, with $\chi$ as the comoving distance and $\hat{n}_0$ as the line-of-sight. A harmonic decomposition under the flat-sky limit can be approximated to be a 2D Fourier transform over $\vec{\theta}$.}
    \label{fig: Flat Sky}
\end{figure}

For a small patch of sky (high multiple moments), flat-sky approximation is sufficiently accurate to calculate the cross-correlation power spectra, for which one can simply replace the spherical harmonic expansion with the 2D Fourier transformation. In this case, the field position can be written as $\vec{x}=\chi \hat{n} \simeq \chi \left(\hat{n}_0 + \hat{\theta}\right)$, where $\hat{n}_0$ is the line-of-sight direction of the field center and $\hat{\theta}$ is the deviation angle in the plane (Fig.~\ref{fig: Flat Sky}). The Fourier modes can be calculated as
\begin{eqnarray}
\label{eq: Xl_definition}
    X(\vec{\ell}) &=& B_X(\ell) \int \md^2\vec{\theta}~ \me^{-i \vec{\ell}\cdot \vec{\theta}} \int\md\chi W_X(\chi) \delta_X(\vec{x}) \nonumber\\
    &=& B_X(\ell) \int\md\chi W_X(\chi) \int \md^2\vec{\theta}~ \me^{-i \vec{\ell}\cdot \vec{\theta}}  \nonumber \\
    &\times & \int \frac{\md^3\vec{k}}{(2\pi)^3}\delta_X(\vec{k}) \me^{i\vec{k}\cdot\vec{x}}.
\end{eqnarray}
For convenience, we choose the coordinate with $\hat{n}_0  = \hat{z}$. Notice that $i\vec{k}\cdot\vec{x} = i k^{\pll} \chi + i \vec{k}^\perp \cdot \chi \vec{\theta}$, the integral over $\vec{\theta}$ can be performed as
\begin{align} \label{eq: delta_D^2 of kperp}
\begin{aligned}
     \int \md^2\vec{\theta} ~\me^{-i\vec{\ell}\cdot \hat{\theta} + i\vec{k}\cdot\vec{x}} &= \me^{i k^{\pll} \chi} \int \md^2\vec{\theta} ~\me^{i\left( \chi\vec{k}^\perp - \vec{\ell} \right) \cdot \vec{\theta}} \\
     &= \me^{i k^{\pll} \chi} \frac{(2\pi)^2}{\chi^2} \delta_{\mathrm{D}}^{(2)}\left(\vec{k}^\perp - \frac{\vec{\ell}}{\chi}\right).
\end{aligned}
\end{align}
Plugging this equation back into Eq.~\eqref{eq: Xl_definition} leads to
\begin{align}\label{eq: Xl}
    X(\vec{\ell}) = B_X(\ell) \int \md\chi \frac{W_X(\chi)}{\chi^2} \int \frac{\md k^{\pll}}{2\pi} \me^{ik^{\pll}\chi} \delta_X(\vec{\ell}/\chi,k^{\pll}).
\end{align}

We now apply this general formula to the kSZ field. Under flat-sky approximation ($\hat{n}\cdot \vec{q} \simeq \hat{n}_0 \cdot \vec{q} \equiv q_z$), it reads
\begin{eqnarray}
\label{eq: deltaT_kSZ of l}
    \delta T_{\mathrm{kSZ}}(\vec{\ell}) &=&  B_{\mathrm{kSZ}}(\ell) \int\md\chi \frac{g(\chi) T_{\mathrm{CMB}}}{\chi^2}  \nonumber \\
&\times &    \int \frac{\md k^{\pll}}{2\pi} \me^{ik^{\pll} \chi} q_z(\vec{\ell}/\chi, k^{\pll}).
\end{eqnarray}
with the Fourier-space momentum function as
\begin{align}\label{eq: q of v and delta}
    \vec{q}(\vec{k}) = \vec{v}(\vec{k}) + \int \frac{\md^3 \vec{p}}{(2\pi)^3} \delta_\mathrm{e}(\vec{k} - \vec{p}) \vec{v}(\vec{p}).
\end{align}
From the linear perturbation theory, the Fourier-space electron's peculiar velocity can be expressed as $\vec{v}(\vec{k})\equiv v(\vec{k})\hat{k}$~\cite{Peebles1980,Ma2011,Ma2012a,Ma2012b,Ma2013a,Ma2013b,Johnson2014,Ma2014a,Ma2014b,Ma2015,Ma2018}, where
\begin{align}
\label{eq: velocity of k}
    v(\vec{k}) = i \frac{aHf \delta_\mathrm{m}(\vec{k})}{ck},
\end{align}
and $f = \md\ln D/\md \ln a$ is the dimensionless linear growth rate for matter perturbations~\cite{Peebles1980,Dodelson2020}.

We can then apply the procedure to the 3D fields of interest. The projected HI intensity mapping field is given by
\begin{align}\label{eq: delta T_HI projection}
    \delta T_\mathrm{HI}(\hat{n}) = B_\mathrm{HI} * \int\md\chi W_{\mathrm{HI}}(\chi) \bar{T}_\mathrm{b}(\chi) \delta_{\mathrm{HI}}(\vec{x}),
\end{align}
where $B_{\mathrm{HI}}$ denotes the beam factor for HI intensity mapping. In real 21-cm experiments, low-$k^{\pll}$ modes are often lost due to the foreground removal (foregrounds are smooth in the $k^{\pll}$ direction), so effectively a filter must be applied to the theory to mimic this observational consequence. In the stage of numerical evaluation, we will apply a low-$k^{\parallel}$ mode cutoff $k^{\pll}_{\mathrm{min}}=0.01~\mathrm{Mpc}^{-1}$ as the effective filter. $\bar{T}_\mathrm{b}(\chi)$ denotes the average brightness temperature at comoving distance $\chi$, which can be evaluated as~\cite{Harper:2017gln,Battye:2012tg}
\begin{align}
    \bar{T}_{\mathrm{b}}(z) = 0.18~\mathrm{K}~\Omega_{\mathrm{HI}} h \frac{(1+z)^2}{E(z)}. \label{eq:Tb_bar}
\end{align}
Here we adopt a recently simulated and widely used value $\Omega_\mathrm{HI}=6.73\times 10^{-4}$~\cite{Jiang:2023zex,Villaescusa-Navarro:2018vsg}, while noticing an early simulation value $\Omega_{\mathrm{HI}} h_{75}=3.5\times 10^{-4}$~\cite{Zwaan:2005cz}.
By using Eq.~\eqref{eq: Xl} we arrive at
\begin{align}
    \delta T_{\mathrm{HI}}(\vec{\ell}) = B_{\mathrm{HI}}(\ell)\int&\md\chi \frac{W_{\mathrm{HI}} 
    (\chi)\bar{T}_\mathrm{b}(\chi)}{\chi^2} \nonumber \\
    &\times\int \frac{\md k^{\pll}}{2\pi} \me^{ik^{\pll}\chi} \delta_{\mathrm{HI}}(\vec{\ell}/\chi,k^{\pll}).
\end{align}
For the projection of squared HI field, simply replacing $\bar{T}_\mathrm{b}(\chi) \delta_{\mathrm{HI}}(\vec{x})$ with its square in Eq.~\eqref{eq: delta T_HI projection} gives
\begin{align}\label{eq: THI^2 of l}
\begin{aligned}
    \delta T_{\mathrm{HI}^2}(\vec{\ell}) = & \int \md\chi ~ \frac{W_{\mathrm{HI}}(\chi)\bar{T}^{2}_\mathrm{b}(\chi)}{\chi^2} \int \frac{\md k^{\pll}}{2\pi} \me^{ik^{\pll}\chi} \\
    &~\times\int\frac{\md^3\vec{p}}{(2\pi)^3} ~ \tilde{\delta}_{\mathrm{HI}} \left( \frac{\vec{\ell}}{\chi} + k^{\pll} \hat{n}_0 - \vec{p}\right) \tilde{\delta}_{\mathrm{HI}}(\vec{p}) ,
\end{aligned}
\end{align}
where we have adopted the following notation for the beam-smoothed 21-cm field
\begin{eqnarray}
    && \tilde{\delta}_{\mathrm{HI}}(\vec{p}) =  B_{\mathrm{HI}}(p^\perp \chi) \delta_{\mathrm{HI}}(\vec{p}), \\
    && \tilde{\delta}_{\mathrm{HI}} \left( \frac{\vec{\ell}}{\chi} + k^{\pll} \hat{n}_0 - \vec{p}\right) =B_{\mathrm{HI}}(|\vec{\ell} - \vec{p}^\perp \chi|)  \nonumber \\
    && \hspace{9em} \times \delta_{\mathrm{HI}} \left( \frac{\vec{\ell}}{\chi} + k^{\pll} \hat{n}_0 - \vec{p}\right),
\end{eqnarray}
where $\vec{p}^\perp = \vec{p}-(\vec{p}\cdot\hat{n}_0)\hat{n}_0$ is the projected wave number.

\subsection{Cross-correlating kSZ square field with projection of the HI square field}
\label{subsec: cross-correlation calculation}
We define the cross-correlation power spectrum of the kSZ square field $\delta T_{\mathrm{kSZ}^2}$ and the projection of the HI intensity mapping square field $\delta T_{\mathrm{HI}^2}$ as
\begin{align} \label{eq: cross-correlation square field}
    \left\langle \delta T_{\mathrm{kSZ}^2}(\vec{\ell}) \delta T_{\mathrm{HI}^2}(\vec{\ell}') \right\rangle = (2\pi)^2 \delta_{\mathrm{D}}^{(2)}(\vec{\ell}+\vec{\ell}') C_\ell^{\mathrm{kSZ}^2\times\mathrm{HI}^2},
\end{align}
where the Fourier transform of the squared-kSZ field follows the convolution theorem as
\begin{align}
    \delta T_{\mathrm{kSZ}^2}(\vec{\ell}) &= \int \frac{\md^2 \vec{\ell}_1}{(2\pi)^2} 
    \delta T_\mathrm{kSZ}(\vec{\ell} - \vec{\ell}_1)
    \delta T_\mathrm{kSZ}(\vec{\ell}_1)~.
\end{align}
Combining $\delta T_{\mathrm{HI}^2}(\vec{\ell})$ in Eq.~\eqref{eq: THI^2 of l}, the angular power spectrum of the kSZ$^{2}$-HI$^{2}$ can be evaluated as
\begin{widetext}
\begin{align} \label{eq: C_l vector definition}
    C(\vec{\ell}) =& \int \frac{\md^2 \vec{\ell}'}{(2\pi)^2} \left\langle \delta T_{\mathrm{kSZ}^2}(\vec{\ell}) \delta T_{\mathrm{HI}^2}(\vec{\ell}') \right\rangle \\
    =& \int \frac{\md^2\vec{\ell}' \md^2\vec{\ell}_1}{(2\pi)^4} \int\md\chi \frac{W_{\mathrm{HI}}(\chi)\bar{T}_\mathrm{b}(\chi)^2}{\chi^2} \int \frac{\md k^{\pll}}{2\pi} \me^{ik^{\pll}\chi}\int\frac{\md^3\vec{p}}{(2\pi)^3} \left\langle  \delta T_\mathrm{kSZ}(\vec{\ell} - \vec{\ell}_1)
    \delta T_\mathrm{kSZ}(\vec{\ell}_1) \tilde{\delta}_{\mathrm{HI}} \left( \frac{\vec{\ell}}{\chi} + k^{\pll} \hat{n}_0 - \vec{p}\right) \tilde{\delta}_{\mathrm{HI}}(\vec{p}) \right\rangle  \nonumber\\
    =& \int \frac{\md^2\vec{\ell}' \md^2\vec{\ell}_1 }{(2\pi)^4} 
    B_{\mathrm{kSZ}}(|\vec{\ell} - \vec{\ell}_1|)
    B_{\mathrm{kSZ}}(\ell_1)
    \int \md\chi_1 \md\chi_2 \md\chi_3  F(\chi_1)F(\chi_2) G(\chi_3) \int\frac{\md k^{\pll}_1 \md k^{\pll}_2 \md k^{\pll}_3}{(2\pi)^3} \exp\left({ i\sum_{n=1}^3 k^{\pll}_n\chi_n }\right) \nonumber\\
    & \quad \times \int\frac{\md^3\vec{p}}{(2\pi)^3}
    B_{\mathrm{HI}}(p^\perp)B_{\mathrm{HI}}(|\vec{\ell}'-\vec{p}^\perp|) \left\langle
    q_z\left(\frac{\vec{\ell} - \vec{\ell}_1}{\chi_1} + k^{\pll}_1 \hat{n}_0\right)
    q_z\left(\frac{\vec{\ell}_1}{\chi_2} + k^{\pll}_2 \hat{n}_0\right)
    \delta_{\mathrm{HI}}\left( \frac{\vec{\ell}'}{\chi_3}  + k^{\pll}_3 \hat{n}_0- \vec{p}\right)
    \delta_{\mathrm{HI}}\left( \vec{p}\right)
    \right\rangle, \nonumber 
\end{align}
with the following abbreviations 
\begin{align}
    F(\chi) \equiv T_\mathrm{CMB} \frac{g(\chi)}{\chi^2}, \quad G(\chi) \equiv \bar{T}_\mathrm{b}(\chi)^2 \frac{W_\mathrm{HI}(\chi)}{\chi^2}~.
\end{align}
The four-point correlation term $\mathcal{T}\equiv \langle q_z q_z \delta \delta \rangle$
should be proportional to a Dirac delta function by momentum conservation/translation invariance, together with a triple spectrum $\tilde{\mathcal{T}}$ factor
\begin{align} \label{eq: qz qz delta delta}
\begin{aligned}
    & ~ \mathcal{T} \left(
    \frac{\vec{\ell} - \vec{\ell}_1}{\chi_1} + k^{\pll}_1 \hat{n}_0,~
    \frac{\vec{\ell}_1}{\chi_2} + k^{\pll}_2 \hat{n}_0,~
     \frac{\vec{\ell}'}{\chi_3}  + k^{\pll}_3 \hat{n}_0- \vec{p},~
    \vec{p} \right) \\
    \equiv &~ \left\langle
    q_z\left(\frac{\vec{\ell} - \vec{\ell}_1}{\chi_1} + k^{\pll}_1 \hat{n}_0\right)
    q_z\left(\frac{\vec{\ell}_1}{\chi_2} + k^{\pll}_2 \hat{n}_0\right)
    \delta_{\mathrm{HI}}\left( \frac{\vec{\ell}'}{\chi_3}  + k^{\pll}_3 \hat{n}_0- \vec{p}\right)
    \delta_{\mathrm{HI}}\left( \vec{p}\right)
    \right\rangle \\
    =&~ (2\pi)^3 \delta_\mathrm{D}^{(2)}\left(
    \frac{\vec{\ell} - \vec{\ell}_1}{\chi_1} +
    \frac{\vec{\ell}_1}{\chi_2} + 
    \frac{\vec{\ell}'}{\chi_3}
    \right)\delta_\mathrm{D}(k^{\pll}_1 + k^{\pll}_2 + k^{\pll}_3)  \tilde{\mathcal{T}} \left(
    \frac{\vec{\ell} - \vec{\ell}_1}{\chi_1} + k^{\pll}_1 \hat{n}_0,~
    \frac{\vec{\ell}_1}{\chi_2} + k^{\pll}_2 \hat{n}_0,~
     \frac{\vec{\ell}'}{\chi_3}  + k^{\pll}_3 \hat{n}_0- \vec{p},~
    \vec{p} \right).
\end{aligned}
\end{align}
\end{widetext}
For small angular scales $k^{\pll}\chi \ll \ell$, one can apply the Limber approximation, allowing us to neglect any $k^{\pll}$ dependence in the triple spectrum as long as the parallel mode is not enrolled in the convolution; therefore the integral over $k^{\pll}_1$, $k^{\pll}_2$ and $k^{\pll}_3$ can be performed to obtain the Dirac delta functions of comoving distances
\begin{align}
\begin{aligned}
    \int\frac{\md k^{\pll}_1 \md k^{\pll}_2 \md k^{\pll}_3}{(2\pi)^3}~ \exp\left({ i\sum_{n=1}^3 k^{\pll}_n\chi_n }\right) (2\pi) \delta_\mathrm{D}(k^{\pll}_1 + k^{\pll}_2 + k^{\pll}_3)~~\\
    = \delta_{\mathrm{D}}(\chi_2-\chi_1)\delta_{\mathrm{D}}(\chi_3-\chi_1)~,
\end{aligned}
\end{align}
which further simplifies the Dirac delta function of perpendicular modes as $\delta_\mathrm{D}^{(2)}\left(\vec{\ell} + \vec{\ell}'\right)$. Integrating over $\vec{\ell}'$ leads to the $C(\vec{\ell})$ expression as follows
\begin{eqnarray} 
\label{eq: C_l vector}
    C(\vec{\ell}) &=& \int\md\chi F(\chi)^2 G(\chi) \int \frac{\md^2\vec{\ell}_1 \md^3\vec{p}}{(2\pi)^5} ~\chi^2 B_{\mathrm{HI}}(p^\perp\chi) \nonumber \\
    & \times & B_{\mathrm{HI}}(|\vec{\ell} + \vec{p}^\perp\chi|) B_{\mathrm{kSZ}}(|\vec{\ell} - \vec{\ell}_1|) B_{\mathrm{kSZ}}(\ell_1) \nonumber \\
    & \times &  \tilde{\mathcal{T}} \left(\frac{\vec{\ell} - \vec{\ell}_1}{\chi},~ \frac{\vec{\ell}_1}{\chi},~ -\frac{\vec{\ell}}{\chi}-\vec{p},~ \vec{p} \right),
\end{eqnarray}
where the $\chi^2$ factor comes from the property of Dirac delta function $\delta_{\mathrm{D}}^{(2)}(\vec{\ell}/\chi) = \chi^{2}\delta_{\mathrm{D}}^{(2)}(\vec{\ell})$.

One remark here is that because $C_\ell$ is only a function of scalar $\ell\equiv |\vec{\ell}|$  due to statistical isotropy, the final expression of the angular power spectrum should be evaluated by performing an angular average of Eq.~\eqref{eq: C_l vector}
\begin{align}\label{eq: C_l average}
    C_\ell^{\mathrm{kSZ}^2\times\mathrm{HI}^2} = \frac{1}{2\pi}\int \md \theta_{\ell}~ C(\vec{\ell})~.
\end{align}
The integrals over $\vec{\ell}_1$ in Eq.~\eqref{eq: C_l vector} are independent of the orientation of $\vec{\ell}$, so $C(\vec{\ell})$ can be evaluated with $\ell$ chosen in an arbitrary direction, for example $\vec{\ell} = \ell\hat{x}$, as the average value. Removing the integral in Eq.~\eqref{eq: C_l average} results in $C_\ell^{\mathrm{kSZ}^2\times\mathrm{HI}^2} = C(\ell\hat{x})$. 

\subsection{Triple-spectrum calculation}\label{subsec: triple-spectrum}

Here we calculate the four-point correlation term appearing in triple-spectrum $\tilde{\mathcal{T}}$, or equivalently the four-point correlator $\mathcal{T}$. 
An important feature of the kSZ effect~\cite{Kaiser:1984,Jaffe:1997ye,Ma:2001xr} is that most contributions to the signal in integral \eqref{eq: deltaT kSZ / TCMB} comes from modes perpendicular to the line-of-sight with $\hat{n}_0\cdot \vec{k}=0$ because of (near) cancellations along the line-of-sight direction. Any 3D vector field can be decomposed into a curl-free component ($E$ field) and divergence-free component ($B$ field), $\vec{q} = \vec{q}_E + \vec{q}_B$, satisfying $\nabla \times \vec{q}_E = 0$ and $\nabla\cdot\vec{q}_B=0$, or in Fourier space $\vec{q}_E\propto\hat{k}$ and $\vec{k}\cdot \vec{q}_B=0$. Then from the definition given in Eq.~\eqref{eq: q of v and delta}, the line-of-sight projection can be evaluated as
\begin{align}\label{eq: qz}
\begin{aligned}
    q_z(\vec{k}) &= \hat{n}_0 \cdot \left(\vec{q}_E(\vec{k}) + \vec{q}_B(\vec{k})\right) \simeq \hat{n}_0 \cdot \vec{q}_B(\vec{k}) \\
    &\simeq \int \frac{\md^3 \vec{p}}{(2\pi)^3}~\left(\hat{p}\cdot\hat{n}_0\right) \delta_\mathrm{e}(\vec{k} - \vec{p}) v(\vec{p})~,
\end{aligned}
\end{align}
where have used the fact that the dominant contribution of the kSZ effect comes from the $B$ field. 

\begin{table*}[ht]
    \centering
    \renewcommand\arraystretch{1.8}
    \begin{tabular}{ccc}
        \hline
        & Expression & Number of identical terms \\
        \hline
        Term 1 & $\quad \left\langle
    \delta_\mathrm{e}(\vec{k}_1 - \vec{p}_1) v(\vec{p}_1) 
    \right\rangle\left\langle
    \delta_\mathrm{e}(\vec{k}_2 - \vec{p}_2) v(\vec{p}_2) 
    \right\rangle\left\langle
    \delta_{\mathrm{HI}} (\vec{k}_3)
    \delta_{\mathrm{HI}} (\vec{k}_4)
    \right\rangle \quad $ & 1 \\
        Term 2 & $\left\langle
    \delta_\mathrm{e}(\vec{k}_1 - \vec{p}_1) 
    \delta_\mathrm{e}(\vec{k}_2 - \vec{p}_2)
    \right\rangle\left\langle
    v(\vec{p}_1)  v(\vec{p}_2) 
    \right\rangle\left\langle
    \delta_{\mathrm{HI}} (\vec{k}_3)
    \delta_{\mathrm{HI}} (\vec{k}_4)
    \right\rangle $ & 1 \\
        Term~3 & $\left\langle
    \delta_\mathrm{e}(\vec{k}_1 - \vec{p}_1) v(\vec{p}_2) 
    \right\rangle\left\langle
    v(\vec{p}_1) \delta_\mathrm{e}(\vec{k}_2 - \vec{p}_2)  
    \right\rangle\left\langle
    \delta_{\mathrm{HI}} (\vec{k}_3)
    \delta_{\mathrm{HI}} (\vec{k}_4)
    \right\rangle$ & 1 \\
    
        Term 4 & $\left\langle
    \delta_\mathrm{e}(\vec{k}_1 - \vec{p}_1) v(\vec{p}_1) 
    \right\rangle\left\langle
    \delta_\mathrm{e}(\vec{k}_2 - \vec{p}_2)
    \delta_{\mathrm{HI}} (\vec{k}_3)
    \right\rangle\left\langle
    v(\vec{p}_2) 
    \delta_{\mathrm{HI}} (\vec{k}_4)
    \right\rangle$ & 2 \\
        Term 5 & $\left\langle
    \delta_\mathrm{e}(\vec{k}_1 - \vec{p}_1) 
    \delta_{\mathrm{HI}} (\vec{k}_3)
    \right\rangle\left\langle
    v(\vec{p}_1) 
    \delta_{\mathrm{HI}} (\vec{k}_4)
    \right\rangle\left\langle
    \delta_\mathrm{e}(\vec{k}_2 - \vec{p}_2) 
    v(\vec{p}_2) 
    \right\rangle$ & 2 \\
        
        Term 6 & $\left\langle
    \delta_\mathrm{e}(\vec{k}_1 - \vec{p}_1) 
    \delta_\mathrm{e}(\vec{k}_2 - \vec{p}_2)
    \right\rangle\left\langle
    v(\vec{p}_1) \delta_{\mathrm{HI}} (\vec{k}_3) 
    \right\rangle\left\langle
     v(\vec{p}_2) \delta_{\mathrm{HI}} (\vec{k}_4)
    \right\rangle$ & 2 \\
        
        Term 7 & $\left\langle
    \delta_\mathrm{e}(\vec{k}_1 - \vec{p}_1) 
    \delta_{\mathrm{HI}} (\vec{k}_3)
    \right\rangle\left\langle
    v(\vec{p}_1) v(\vec{p}_2)
    \right\rangle\left\langle
    \delta_\mathrm{e}(\vec{k}_2 - \vec{p}_2) 
    \delta_{\mathrm{HI}} (\vec{k}_4)
    \right\rangle$ & 2 \\
        
        Term 8 & $\left\langle
    \delta_\mathrm{e}(\vec{k}_1 - \vec{p}_1) v(\vec{p}_2) 
    \right\rangle\left\langle
    \delta_\mathrm{e}(\vec{k}_2 - \vec{p}_2)
    \delta_{\mathrm{HI}} (\vec{k}_3)
    \right\rangle\left\langle
    v(\vec{p}_1) \delta_{\mathrm{HI}} (\vec{k}_4)
    \right\rangle$ & 4 \\
        \hline
        Total & & 15 \\
        \hline
    \end{tabular}
    \caption{Independent terms and the corresponding numbers of identical terms appearing in Wick expansion.}
    \label{tab:Wick Terms}
\end{table*}

The four-point correlator can be evaluated by substituting Eq.~\eqref{eq: qz} into Eq.~\eqref{eq: qz qz delta delta},
\begin{widetext}
\begin{align}\label{eq:definition of T}
\begin{aligned}
    \mathcal{T}(\vec{k}_1,\vec{k}_2,\vec{k}_3,\vec{k}_4) = \int \frac{\md^3 \vec{p}_1 \md^3 \vec{p}_2}{(2\pi)^6} \mu_1 \mu_2 
    \left\langle 
    \delta_\mathrm{e}(\vec{k}_1 - \vec{p}_1) v(\vec{p}_1) 
    \delta_\mathrm{e}(\vec{k}_2 - \vec{p}_2) v(\vec{p}_2) 
    \delta_{\mathrm{HI}} (\vec{k}_3)
    \delta_{\mathrm{HI}} (\vec{k}_4)
    \right\rangle,
\end{aligned}
\end{align}
\end{widetext}
where $\mu_1 = \hat{n}_0 \cdot \hat{p}_1$ and $\mu_2 = \hat{n}_0 \cdot \hat{p}_2$. Further assuming Gaussianity of the perturbation fields allows us to expand the six-point correlation with the Wick Theorem, to a summation of 15 terms in total (see, e.g. Table~\ref{tab:Wick Terms}). Before we move into the details of calculation, by observing the underlying symmetry of the arguments in $\mathcal{T}$, some simplifications and reduction can be carried out. First, permutation of $\vec{k}_3$ and $\vec{k}_4$ does not change the integral, and therefore some of the terms give exactly the same contribution, at the integrand level. For the same reason, permutation of the vector pair $(\vec{k}_1,\vec{p}_1)$ and $(\vec{k_2},\vec{p}_2)$ will also result in the same integral because of the exchangeability between $\vec{k}_1$ and $\vec{k}_2$. As a consequence, all independent terms are shown in Table~\ref{tab:Wick Terms} explicitly, where the power spectrum of any two-point correlation is given by
\begin{align}
    \left\langle X(\vec{k}) Y(\vec{k}') \right\rangle = (2\pi)^3 \delta_\mathrm{D}^{(3)}(\vec{k} + \vec{k}') P_{X,Y}(k)
\end{align}
with $X,Y = \mathrm{e}, v, \mathrm{HI}$ and $P_{X,Y}$ is the cross-correlation power spectrum. We should also keep in mind of doing the following replacement in the Fourier modes at the final step
\begin{align}\label{eq: k_replacement}
    \begin{aligned}
        \vec{k}_1 \to \frac{\vec{\ell} - \vec{\ell}_1}{\chi}~,\quad
        \vec{k}_2 \to \frac{\vec{\ell}_1}{\chi}~, \quad
        \vec{k}_3 \to -\frac{\vec{\ell}}{\chi}-\vec{p}~,\quad
        \vec{k}_4 \to \vec{p}~.
    \end{aligned}
\end{align}

We are now ready to calculate each individual term. Fortunately, it is not too difficult to see the first five terms in Table~\ref{tab:Wick Terms} are all equal to zero by using simple arguments. The first term is evaluated as
\begin{align}
\begin{aligned}
    \mathcal{T}_1 &= \int \frac{\md^3 \vec{p}_1 \md^3 \vec{p}_2}{(2\pi)^6} \mu_1 \mu_2 (\mathrm{Term}~1)\\ &\propto \delta_\mathrm{D}^{(3)}(\vec{k}_1)
    \delta_\mathrm{D}^{(3)}(\vec{k}_2)\delta_\mathrm{D}^{(3)}(\vec{k}_3 + \vec{k}_4).
\end{aligned}
\end{align}
By applying the arguments replacement Eq.\eqref{eq: k_replacement}, the Dirac delta function $\delta_\mathrm{D}^{(3)}(\vec{k}_3 + \vec{k}_4) \to \delta_\mathrm{D}(k_3^{\pll}) \delta_\mathrm{D}^{(2)}(\vec{\ell}'/\chi_3)$ indicates a zero contribution to $C_\ell$ because it gives out a factor proportional to $\delta_\mathrm{D}^{(2)}(\vec{\ell}')$. The cross-correlation Eq.~\eqref{eq: cross-correlation square field} has put a constraint $\vec{\ell} + \vec{\ell}' = 0$ on $\ell$, suggesting that Term $1$ will only contribute to the monopole $\ell=0$, which is not considered. The same factor $\delta_\mathrm{D}^{(3)}(\vec{k}_3 + \vec{k}_4)$ appears in Terms $2$ and $3$, and thus there is no contribution from these two terms either. Next is Term $4$:
\begin{align}
\begin{aligned}
    \mathcal{T}_4 &= \int \frac{\md^3 \vec{p}_1 \md^3 \vec{p}_2}{(2\pi)^6} \mu_1 \mu_2 (\mathrm{Term}~4)\\ 
    &\propto \int_0^{\infty} \frac{\md p_1}{2\pi} \left(\frac{p_1}{2\pi}\right)^2 P_{\mathrm{e},v}(p_1) \int_0^{2\pi}\md\phi\int_{-1}^{1}\mu_1\md\mu_1  = 0.
\end{aligned}
\end{align}
Mathematically, the resultant zero value in the above evaluation is because Term $2$ is proportional to $\delta_\mathrm{D}^{(3)}(\vec{k}_1)$, which does not pick up a specific value of $\vec{p}_1$ in the integral, but rather sums over all the directions, leading to a factor proportional to zero. Similarly, Term $5$ is also zero because it is proportional to $\delta_\mathrm{D}^{(3)}(\vec{k}_2)$.

\begin{widetext}
Therefore, the first {\it nontrivial} contribution is from Term $6$
\begin{align}\label{eq: T_6 contribution}
\begin{aligned}
    \mathcal{T}_6 =& \int \frac{\md^3 \vec{p}_1 \md^3 \vec{p}_2}{(2\pi)^6} \mu_1 \mu_2  (\mathrm{Term}~6) = -(2\pi)^3 \delta_\mathrm{D}^{(3)}(\vec{k}_1 + \vec{k}_2 + \vec{k}_3 + \vec{k}_4) \frac{k_3^{\pll}}{k_3} \frac{k_4^{\pll}}{k_4}  P_{\mathrm{e},\mathrm{e}}(|\vec{k}_1 + \vec{k}_3|)  P_{v,\mathrm{HI}} (k_3) P_{v,\mathrm{HI}} (k_4) ,
\end{aligned}
\end{align}
where the minus sign in the front results from two copies of velocity fields with imaginary parts, as shown in Eq.~\eqref{eq: velocity of k}. Note that $k^{\pll}_3$ does not stand for the $k^{\pll}_3$ in Eq.~\eqref{eq: C_l vector definition} but the $z$-component of $\vec{k}_3$. Applying the replacement \eqref{eq: k_replacement}, the triple-spectrum from Term $6$ reads
\begin{align}
    \mathcal{T}_6 \to \frac{(p^{\pll}\chi)^2}{|\vec{\ell}+\vec{p}\chi|\cdot p\chi} P_{\mathrm{e},\mathrm{e}}\left(\left| \frac{\vec{\ell}_1}{\chi} + \vec{p} \right|\right) P_{v,\mathrm{HI}}\left(\left|\frac{\vec{\ell}}{\chi} + \vec{p}\right|\right) P_{v,\mathrm{HI}} \left( p \right).
\end{align}
Assembling all the delta function and evaluating integrals in Eq.~\eqref{eq: C_l vector} leads to the final result for contributions to cross-correlation from Term $6$
\begin{align}
    \begin{aligned}
        C_\ell^{(6)} =& \int\md\chi F(\chi)^2 G(\chi) \int\frac{\md^2 \vec{\ell}_1 \md^3 \vec{p}}{(2\pi)^5} \chi^2~B_{\mathrm{kSZ}}(|\vec{\ell} - \vec{\ell}_1|) B_{\mathrm{kSZ}}(\ell_1) B_{\mathrm{HI}}(|\vec{\ell} + \vec{p}^\perp\chi|)B_{\mathrm{HI}}(p^\perp\chi) \\
        &\quad\quad \times \frac{(p^{\pll}\chi)^2}{|\vec{\ell} + \vec{p}\chi|\cdot p\chi}
        ~P_{\mathrm{e},\mathrm{e}}\left(\left| \frac{\vec{\ell}_1}{\chi} + \vec{p} \right|\right) P_{v,\mathrm{HI}}\left(\left|\frac{\vec{\ell}}{\chi} + \vec{p}\right|\right) P_{v,\mathrm{HI}} \left( p \right)~.
    \end{aligned}
\end{align}
One can perform a decomposition of convolution momentum $\vec{p} = \vec{p}^\perp + p^{\pll}\hat{n}_0$ with $p = \sqrt{|\vec{p}^\perp|^2 + (p^{\pll})^2}$, suggesting a 2D integral over $\vec{p}^\perp$ perpendicular to line-of-sight $\hat{n}_0$ and a 1D integral along $\hat{n}_0$. Physically speaking, the integral over $p^{\pll}$ carries the information about the parallel modes of the HI field, which is absent in the kSZ square field, because the kSZ effect is already projected to 2D before taking the square.

Similar to Term 6, we can write down all the contributions in the following form
\begin{align}\label{eq: Cli}
    C_\ell^{(i)} = \int\md\chi F(\chi)^2 G(\chi) \int\frac{\md^2 \vec{\ell}_1 \md^3 \vec{p}}{(2\pi)^5}\chi^2 B_{\mathrm{kSZ}}(|\vec{\ell} - \vec{\ell}_1|) B_{\mathrm{kSZ}}(\ell_1) B_{\mathrm{HI}}(|\vec{\ell} + \vec{p}^\perp\chi|)B_{\mathrm{HI}}(p^\perp\chi) ~ Y^{(i)},
\end{align}
with $Y^{(i)}$ defined as three independent nonvanishing contributions from Term $i$ given below
\begin{eqnarray}
\label{eq: Yi results}
    Y^{(6)} &= & + \frac{(p^{\pll}\chi)^2}{|\vec{\ell} + \vec{p}\chi|\cdot p\chi} ~
    P_{\mathrm{e},\mathrm{e}}\left(\left| \frac{\vec{\ell}_1}{\chi} + \vec{p} \right|\right) 
    P_{v,\mathrm{HI}}\left(\left|\frac{\vec{\ell}}{\chi} + \vec{p}\right|\right) 
    P_{v,\mathrm{HI}} \left(p\right) , \nonumber 
    \\
    Y^{(7)} &= & + \frac{(p^{\pll}\chi)^2}{|\vec{\ell}_1 + \vec{p}\chi|^2}
    P_{\mathrm{e},\mathrm{HI}}\left(\left|
    \frac{\vec{\ell}}{\chi} + \vec{p}
    \right|\right) 
    P_{v,v}\left(\left|\frac{ \vec{\ell}_1}{\chi}  + \vec{p} \right|\right) 
    P_{\mathrm{e},\mathrm{HI}} \left(p\right), \nonumber 
    \\
    Y^{(8)} &=& - \frac{(p^{\pll}\chi)^2}{|\vec{\ell} - \vec{\ell}_1 + \vec{p}\chi|\cdot p\chi} ~
    P_{\mathrm{e},v}\left(\left| \frac{\vec{\ell} - \vec{\ell}_1}{\chi}+ \vec{p}\right|\right) 
    P_{\mathrm{e},\mathrm{HI}}\left(\left|\frac{\vec{\ell}}{\chi} + \vec{p}\right|\right) 
    P_{v,\mathrm{HI}} \left(p\right).
\end{eqnarray}
\end{widetext}
The integral along parallel modes $p^{\pll}$ appears in the cross-correlation through the convolution when calculating the 2D Fourier transformation of the projected HI square field \eqref{eq: THI^2 of l}, which contains the line-of-sight information that is absent in a pure projected field. Here we have applied a cutoff $p^{\pll}_{\mathrm{min}}=0.01~\mathrm{Mpc}^{-1}$ accounting for the signal loss due to foregrounds, by which the largest line-of-sight scale mode is removed. Although the cross-correlations with the cutoff value above are almost the same comparing with that with a zero cutoff, it should be reminded that a more aggressive $p^{\pll}_{\mathrm{min}}$ (such as $0.1~\mathrm{Mpc}^{-1}$) would further suppress the large-scale signal and diminish the expected signal-to-noise ratio.

Further enrolled in the other convolutions while evaluating the kSZ effects in Eq.~\eqref{eq: qz}, the parallel modes $p^{\pll}$ dependence finally enter all the power spectrum given in Eq.~\eqref{eq: Yi results}.

The complete cross-correlation is obtained by summing over all contributing terms,
\begin{eqnarray}
    C_\ell^{\mathrm{kSZ}^2\times\mathrm{HI}^2} &=& \sum_i  C_\ell^{(i)} \times (\mathrm{Number~of~identical~terms})_i \nonumber \\ 
    &=& 2 C_\ell^{(6)} + 2C_\ell^{(7)} + 4 C_\ell^{(8)},  \label{eq:terms6-7-8}
\end{eqnarray}
where each term carries a different physical meaning. 
$C_\ell^{(6)}$ is proportional to the cross-correlation between the HI field and the velocity field multiplied by the autocorrelation of the electron density perturbation. 
Similarly, $C_\ell^{(7)}$ encodes the correlation between HI and the electron, leaving the bulk velocity to be autocorrelated. These two terms correspond to the cross-correlation between HI and the two kinds of perturbation field components of the kSZ, i.e. velocity and electron distribution, respectively, and both provide a positive contribution to the final cross-correlation. In comparison, $C_\ell^{(8)}$ contains the cross-correlation of three pairs of perturbation fields: electron-velocity, electron-HI, and HI-velocity, with negative contributions to the final cross-correlation.

\subsection{Redshift-space distortion (RSD) corrections}

Since HI intensity mapping is naturally observed in the redshift space, the observed HI intensity has the Doppler effect due to the peculiar velocity of the sources. Therefore, the leading-order redshift-space HI density contrast is
\begin{align}
    \delta_{s,\mathrm{HI}}(\vec{k}) = \left(b_{\mathrm{HI}} + f\mu^2 \right)\delta_\mathrm{m}(\vec{k}) ~,
\end{align}
where $\mu = k^{\pll}/k = \hat{n}\cdot\hat{k}$, $b_\mathrm{HI}$ is the bias for the HI field that relates the HI field with matter perturbation, and the subscript $s$ denotes for the observable in redshift space. Here we put the detailed calculation in Appendix~\ref{app: RSD}~\footnote{For a more complete treatment of linear and nonlinear RSD effects, please refer to Ref.~\cite{Shaw:2008aa}.}

Finally, we need to introduce linear bias factors to relate the perturbations in the electron density, HI distribution, and peculiar velocity field to the underlying matter density fluctuations. 
Since the main signal of interest comes from large scales, the linear biasing approximation should be accurate, and we can take the electron and HI bias factors to be constants, denoted by $b_\mathrm{e}$ and $b_{\mathrm{HI}}$, respectively.
One can also effectively define a ``bias'' for electron velocity field as $b_v \equiv aHf / (ck)$ to rewrite all the cross power spectrum in Eq.~\eqref{eq: Yi results} with matter power spectrum
\begin{align}
    & P_{X,Y}(k) = \tilde{b}_X \tilde{b}_Y P_{\mathrm{m}}(k), \quad X,Y=\mathrm{e},\mathrm{HI},v \\
    & \tilde{b}_\mathrm{e}=b_\mathrm{e},\quad \tilde{b}_{\mathrm{HI}}=b_\mathrm{HI}+f\mu^2 , \quad \tilde{b}_v = \frac{aHf}{ck}~.
\end{align}
Each term in Eq.~\eqref{eq: Yi results} contains two $\delta_\mathrm{e}$, two $v$ and two $\delta_{\mathrm{HI}}$. Then, the cross-correlation power spectrum $C_\ell$ can be simplified to
\begin{widetext}
\begin{align}\label{eq: C_l simplified}
\begin{aligned}
    C_\ell^{\mathrm{kSZ}^2\times\mathrm{HI}^2} = & 2\int\md\chi F(\chi)^2 G(\chi) \left(\frac{aHf}{c}\right)^2    
    \int \frac{\md^2\vec{\ell}_1 \md^3\vec{p}}{(2\pi)^5} \chi^4~\tilde{f}_{\mathrm{kSZ}}(|\vec{\ell} - \vec{\ell}_1|) \tilde{f}_{\mathrm{kSZ}}(\ell_1) B_{\mathrm{HI}}(|\vec{\ell} + \vec{p}^\perp\chi|)B_{\mathrm{HI}}(p^\perp\chi)
    \\
    &\quad\quad\times (p^{\pll}\chi)^2 b_\mathrm{e}^2 b_{\mathrm{HI}}^2  
    P_{\mathrm{m}}\left(\left|\frac{\vec{\ell}}{\chi} + \vec{p}\right|\right)  P_{\mathrm{m}}\left(\left| \frac{\vec{\ell}_1}{\chi} + \vec{p} \right|\right) P_{\mathrm{m}}\left(p\right) ~\mathcal{T}(\vec{\ell},\vec{\ell}_1,\vec{p})~,
\end{aligned}
\end{align}
where $\mathcal{T}$ reads
\begin{align} \label{eq:geo_T}
    \mathcal{T}(\vec{\ell},\vec{\ell}_1,\vec{p}) = 
    \left(\frac{1}{|\vec{\ell}+\vec{p}\chi|^2(p\chi)^2} + \frac{1}{|\vec{\ell}_1+\vec{p}\chi|^4} - \frac{2}{|\vec{\ell}+\vec{p}\chi|^2|\vec{\ell}_1+\vec{p}\chi|^2} \right) \left(1+b_\mathrm{HI}^{-1}f\frac{(p^{\pll})^2}{|\vec{\ell}/\chi+\vec{p}|^2}\right) \left(1+b_\mathrm{HI}^{-1}f\frac{(p^{\pll})^2}{p^2}\right),
\end{align}
\end{widetext}
in which the last two factors account for the RSD effect. The three terms in the first factor correspond to contributions from Terms $6$, $7$ and $8$ respectively. We will show in later numerical results that, there are hierarchies among these three terms because of the differences in geometric factors in Eq.~\eqref{eq:definition of T}, which originate from the direction of velocity perturbation in Fourier space $\hat{v}(\vec{k})=\hat{k}$. Notice that here we replace the kSZ beam with a normalized filter $\tilde{f}_{\mathrm{kSZ}}$ because the kSZ map is obtained from a filtered CMB map. The detailed form of $\tilde{f}_{\mathrm{kSZ}}$ is shown in the next section.

\section{Numerical Evaluations}\label{sec:num_res}

\subsection{Cross-correlation}\label{subsec: num cross}
We now numerically evaluate the cross-correlation function by specifying the kernel functions, beam functions and biases in Eq.~\eqref{eq: Cli}. First, as is given in Eq.~\eqref{eq: g(z)}, the kernel function of the kSZ field is just the visibility function $g(\chi)$, which can be specified for a given cosmological and astrophysical model. The kernel of the HI field is given by the window function of observation
\begin{eqnarray}
&& W_{\mathrm{HI}}(\chi) = \frac{\md z}{\md \chi} W_{\mathrm{HI}}(z)  \nonumber \\
&& = \frac{H(z)}{c}\times \left\{
    \begin{array}{ll}
        1/\Delta z & ,~z_{\mathrm{min}}<z<z_{\mathrm{max}}  \\
        0 & ,~\mathrm{otherwise}
    \end{array}
    \right.  \label{eq:W_HI}
\end{eqnarray}
where $\Delta z=z_{\mathrm{max}}-z_{\mathrm{min}}$, and $z_{\mathrm{min}}$ and $z_{\mathrm{max}}$ are the minimum and maximum redshift coverage of an observation. Based on HI observation by SKA-MID Wide Band 1 Survey \cite{SKA:2018ckk} with approximately $20,000~\mathrm{deg}^2$ sky coverage and $t_{\mathrm{tot}}=10,000$ hours total integration time, we choose $z_{\mathrm{min}}=0.3$ and $z_{\mathrm{max}}=1.0$ as an estimation.\footnote{In fact, SKA-MID Wide Band 1 Survey planned to proceed a HI intensity mapping in a redshift range $z=0.35\sim 3$, which is wider than $z=0.3\sim 1$ we selected. Our choice allows us to perform the analysis within a region where the biases can be approximated as constant, therefore the redshift dependence of the HI bias can be neglected.}

\begin{table}[t]
    \centering
    \begin{tabular}{ccccc}
        \hline
        \multirow{2}{*}{} 
        & Channel & Beam & $\delta T_{\rm N}$ & Depth \\
        & [GHz] & [arcmin] & [$\mu$K-arcmin] & [mJy] \\
        \hline
        \multirow{3}{*}{ACT} 
        & 90  & 1.42 & 14 & 6.5 \\
        & 150 & 2.07 & 14 & 8.4 \\
        & 220 & 1.01 & 64 & 29  \\
        \hline 
        \multirow{6}{*}{SO} 
        & 27  & 7.4 & 61  & 27  \\
        & 39  & 5.1 & 30  & 19  \\
        & 93  & 2.2 & 5.3 & 6.9 \\
        & 145 & 1.4 & 6.6 & 8.3 \\
        & 225 & 1.0 & 15  & 17  \\
        & 280 & 0.9 & 35  & 34  \\
        \hline
    \end{tabular}
    \caption{Instrumental parameters of the CMB experiments for evaluating internal linear combination (ILC) noise power. $\delta T_{\rm N}$ is the {\it rms} noise in unit of $\mu$-arcmin, which is related to the $\Delta_{T}$ quantity by $\Delta_{T}=\pi\delta T_{\rm N}/(180\times 60)$.}
    \label{tab:parameters}
\end{table}
Second, the beam functions are modeled as follows
\begin{align}
    B_{X}(\ell) = \exp\left(-\frac{\ell^2}{2}\frac{\theta_{\mathrm{FWHM}}^{2}}{8\ln2}\right),
\end{align}
where $\theta_{\mathrm{FWHM}}$ is the full-width at half-maximum (FWHM), with $X$ can be either HI or kSZ field. For HI detection, the FWHM is frequency dependent, which is 
\begin{align}
    \theta_{\mathrm{FWHM}}^{\mathrm{HI}} = 1.02 \frac{\lambda}{D_{\mathrm{dish}}} \simeq 0.0227 \left(\frac{\nu}{1000~\mathrm{MHz}}\right)^{-1},
\end{align}
where we have substituted the diameter of MeerKAT single dish $D_{\mathrm{dish}}=13.5$\,m, and $\nu = 1420\,{\rm MHz}/(1+z)$ is the observed frequency of the $21$-cm signal at redshift $z$. One should notice, that the complete SKA-MID dish should have 133 15-meter dishes, apart from the MeerKAT dishes~\cite{SKA:2018ckk}. But here we neglect the difference. For a typical redshift $z=0.5$, such a beam will lead to a $1/\mathrm{e}$ suppression factor at $\ell\simeq 110$.  

For the kSZ effect measurements, the instrumental parameters ACT indicate an FWHM of order $\mathcal{O}(1)$ arcmin for different frequency bands~\cite{Swetz:2010fy}. Similarly, SO, as a pathfinder of CMB-S4 observation, has a similar resolution $\theta_{\mathrm{FWHM}}=1.4'$ ($\ell_\mathrm{max}\simeq8000$). Thus for a simple estimation we adopt the following uniform beam suitable for both ACT and SO
\begin{align}
    \theta_{\mathrm{FWHM}}^{\mathrm{kSZ}} = 1.4' \simeq 2.9\times10^{-4}~\mathrm{radians}.
\end{align}

Besides the beam, there is an additional factor to be applied to the observed CMB map, which is the $\ell$-space filter that suppress the primary CMB and noise components. We apply a Wiener filter~\footnote{In principle, a Wiener filter should also be applied on HI data, which is neglected because we're considering a low-noise future measurement in this work.} $f(\ell)$, together with the beam $B(\ell)$, on the temperature map in harmonic space (see also, e.g., Refs.~\cite{Hill:2016dta,Ferraro:2016ymw})
\begin{align}
\begin{aligned}
   \Theta_f &= f(\ell)B(\ell)\left(\Theta_\mathrm{kSZ} + \Theta_\mathrm{CMB} + \Theta_{\mathrm{noise}}\right) \\
    &\equiv \tilde{f}(\ell) \left(\Theta_\mathrm{kSZ} + \Theta_\mathrm{CMB} + \Theta_{\mathrm{noise}}\right) ,
\end{aligned} 
\end{align}
where we have defined the normalized filter $\tilde{f}(\ell) = f(\ell)B(\ell)$, and the Wiener filter is given by
\begin{align}
    \quad f(\ell) = \frac{C_\ell^\mathrm{kSZ}}{C_\ell^\mathrm{tot}} = \frac{C_\ell^\mathrm{kSZ}}{C_\ell^\mathrm{kSZ} + C_\ell^\mathrm{CMB} + N_\ell}.
\end{align}
The total angular power spectrum $\tilde{C}_\ell^{\mathrm{tot}} \equiv  C_\ell^{\mathrm{CMB}} + C_\ell^{\mathrm{kSZ}} + N_\ell$ contains contributions from primary lensed CMB temperature power spectrum $C_\ell^{\mathrm{CMB}}$, kSZ autocorrelation $C_\ell^{\mathrm{kSZ}}$, and noise angular power $N_\ell$. For a simple estimation, we only consider the instrumental noise with negligible foreground contamination, which is $N_\ell \simeq \Delta_T^2 B(\ell)^{-2}$, with $\Delta_T^2$ being the {\it rms} noise per radian. A more realistic noise model accounts for residual foreground contamination, which can be estimated by evaluating the post-ILC (internal linear combination) noise that is constructed from a weighted linear combination of maps at different frequencies. We use the public package {\sc orphics}\footnote{\hyperlink{https://github.com/msyriac/orphics}{github.com/msyriac/orphics}} code and apply the ILC method in harmonic space to obtain the noise power $N_\ell$ for ACT and SO respectively, whose instrument parameters are shown in Table~\ref{tab:parameters} (see also ACT DR6 data release~\cite{ACT:2025xdm} and SO forecast~\cite{SimonsObservatory:2025wwn}). 
Note that although we apply a Wiener filter to help extract the kSZ signal, our statistic may still receive contributions from CMB lensing~\cite{Ferraro:2016ymw}. In future work, it may be interesting to quantify the effects of CMB lensing contamination on our estimates.

\begin{figure}[t]
    \centering
    \includegraphics[width=0.45\textwidth]{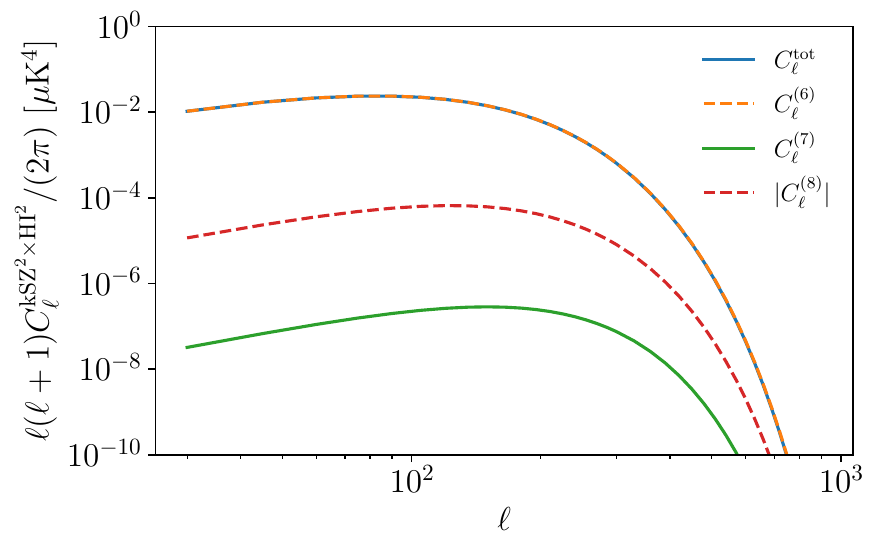}
    \caption{The cross-correlation angular power spectrum with each individual term's contribution Eq.~\eqref{eq:terms6-7-8}, where the biases $b_\mathrm{e}$ and $b_\mathrm{HI}$ are set to be unity. 
    }
    \label{fig:cross-correlation}
\end{figure}

We finally substitute all the instrument parameters into Eq.~\eqref{eq: C_l simplified}, perform the integral numerically and obtain the angular power spectrum. The different contributions received from $C^{(6)}_{\ell}$, $C^{(7)}_{\ell}$ and $C^{(8)}_{\ell}$ and the sum are shown in Fig.~\ref{fig:cross-correlation}. Here the biases $b_\mathrm{e}$ and $b_\mathrm{HI}$ are set to be unity, and the contributions from each term are approximately proportional to the same constant factor $\eta=b_\mathrm{e}^2b_\mathrm{HI}^2\Omega_\mathrm{HI}^2$.\footnote{Despite the nonlinear effects, $C_\ell$ is not precisely proportional to $\eta$ because there are biases dependence in the RSD corrections.} Therefore the total cross-correlation follows a simple relationship proportional to $\eta$.

{Here we briefly discuss each term separately. As we can see, $C^{(6)}_{\ell}$ (HI-velocity cross-correlation) makes up the main contribution to $C_\ell^{\mathrm{kSZ}^2\times\mathrm{HI}^2}$, while $C^{(7)}_{\ell}$ (HI-electron) contributes least. Term 8, although negative, contributes in between. Intuitively speaking, Term 7 contains the autocorrelation of the velocity fields, which should dominate a certain scale, which is not obvious in our result because of a suppression from the HI beam at large $\ell$s. For numerical calculation and analytical approximation, it can be shown that Term 7 is most sensitive to the HI beam effect, followed by Term 8, and finally Term 6, which contributes most under our beam choice. Further discussions for possible variations are conducted in Appendix~\ref{app: power no beam effect}.

\subsection{SNR forecasts}\label{subsec: SNR}

\begin{figure*}[t]
\centerline{\includegraphics[width=0.45\linewidth]{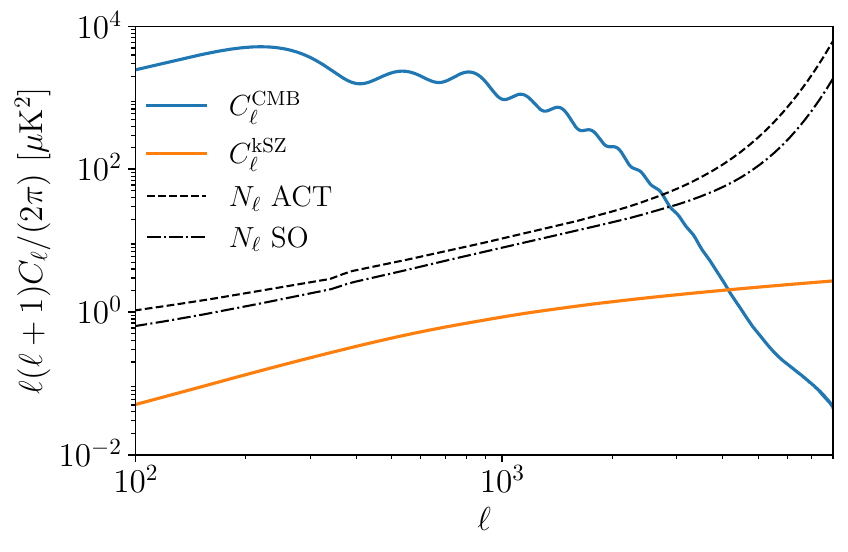}
\includegraphics[width=0.46\linewidth]{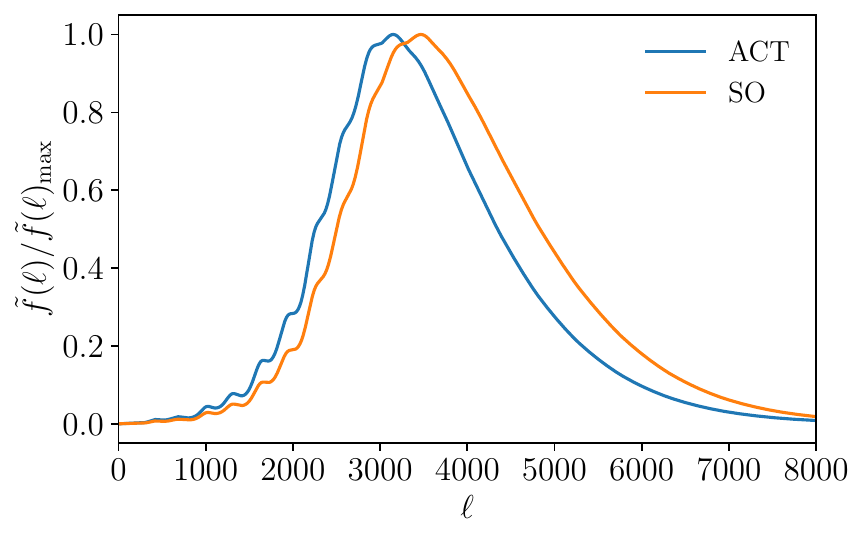}}
\caption{{\it Left}: The angular power spectrum for primary (lensed) CMB (blue solid line), kSZ effect (orange solid line) and the (deconvolved) noise for the ACT telescope (dashed line) and Simons Observatory (SO; dot-dashed line). {\it Right}: The normalized filter for ACT (blue solid line) and SO (orange solid line) respectively.
}
\label{fig:kSZ and filter}
\end{figure*}

The maximum signal-to-noise ratio (SNR) can be estimated by using Fishers' formula \cite{Ferraro:2016ymw}
\begin{align}\label{eq:SNR2}
\begin{aligned}
    \mathrm{SNR}^2 &\equiv \sum_{\ell=\ell_{\mathrm{min}}}^{\ell_{\mathrm{max}}} \mathrm{SNR}_\ell^2 \\    
    &= f_{\mathrm{sky}} \sum_{\ell=\ell_{\mathrm{min}}}^{\ell_{\mathrm{max}}} \frac{(2\ell+1) \left( C_\ell^{\mathrm{kSZ}^2\times\mathrm{HI}^2} \right)^2}{\tilde{C}_\ell^{\mathrm{kSZ}^2} \tilde{C}_\ell^{\mathrm{HI}^2} + \left( C_\ell^{\mathrm{kSZ}^2\times\mathrm{HI}^2} \right)^2}, 
\end{aligned}
\end{align}
where $f_{\mathrm{sky}}$ is the observed sky fraction and the tilded quantities denote the tracers' autocorrelations with noises. For $\tilde{C}_\ell^{\mathrm{kSZ}^2}$, we adopt Gaussian approximation~\cite{Ma:2001xr}
\begin{eqnarray}
    \tilde{C}_\ell^{\mathrm{kSZ}^2} = 2\int \frac{\md^2 \vec{\ell}'}{(2\pi)^2} ~\tilde{C}_{\ell'}^{\mathrm{kSZ}} \tilde{C}_{|\vec{\ell} - \vec{\ell}'|}^{\mathrm{kSZ}}, \label{eq:Cell_kSZ2}
\end{eqnarray}
where $\tilde{C}_\ell^{\mathrm{kSZ}} = f^{2}(\ell)B^{2}(\ell) C_\ell^{\mathrm{tot}}$ is the kSZ-filtered CMB power spectrum. The angular power spectrum for the kSZ projection field (without beam function) is given by~Refs.\cite{Ma:2001xr,Ma2014}
\begin{eqnarray}
    C_\ell^{\mathrm{kSZ}} &=& \int\md\chi \left(\frac{g T_{\mathrm{CMB}}}{\chi}\right)^2\frac{(aHf/c)^2}{2}\int \frac{\md^3 \vec{k}}{(2\pi)^3} \nonumber \\
    &\times& \frac{\ell(\ell - 2k\chi \mu)(1-\mu^2)}{k^2(\ell^2 + (k\chi)^2 - 2k\chi \ell\mu)} P_{\rm e}\left(\left|\frac{\vec{\ell}}{\chi} - \vec{k}\right|\right) P_{\rm e}(k) ~, \nonumber \\
    \label{eq: C_l kSZ}
\end{eqnarray}
where $\mu = \hat{\ell}\cdot\hat{k}$ and $g=g(\chi)$ are defined in Eq.~\eqref{eq: g(z)}. Note here without cross-correlation with HI, the integration of $\chi$ should go all the way to the comoving distance of the last scattering surface ($z_{\ast}\simeq 1100$). A good approximation is to take $\chi_{\mathrm{max}} = \chi(z=10)$, beyond which all atoms in cosmic fluid can be safely assumed to be nonionized with $X_{\rm e}=0$. It should be noted that the early-time kSZ effect (patchy reionization) would have no cross-correlation with late time HI distribution at $z\leq 1$, but can contribute to the kSZ autocorrelation. 
The numerical results of $C_{\ell}^{\mathrm{kSZ}}$, $C_{\ell}^{\mathrm{CMB}}$ (primary CMB) and noise level $N_\ell$ are show in the left panel of Fig.~\ref{fig:kSZ and filter}, and the normalized filters are shown in the right panel of Fig.~\ref{fig:kSZ and filter}. One can see from the left panel that primary CMB dominates the map from very large scales to $\ell\simeq 4000$, after which the kSZ becomes dominant. In addition, because of the lower noise level of SO (left panel), the central kernel of the filter peaks at smaller angular scales comparing to ACT (right panel of Fig.~\ref{fig:kSZ and filter}).

\begin{figure*}[t]
    \centerline{\includegraphics[height=0.3\linewidth]{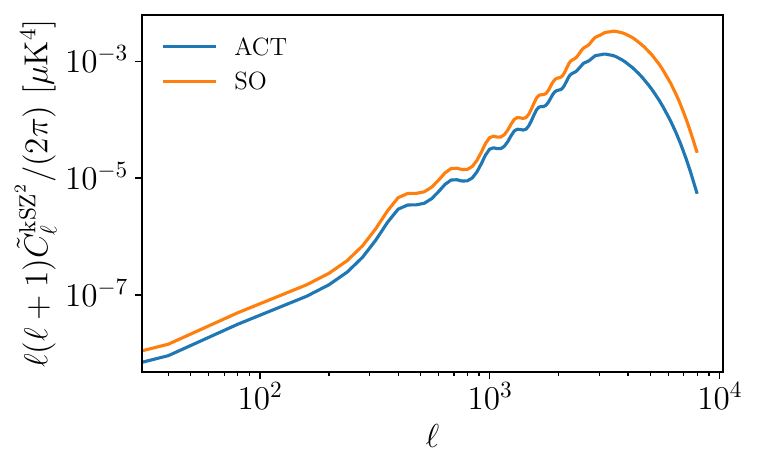}
    \includegraphics[height=0.31\linewidth]{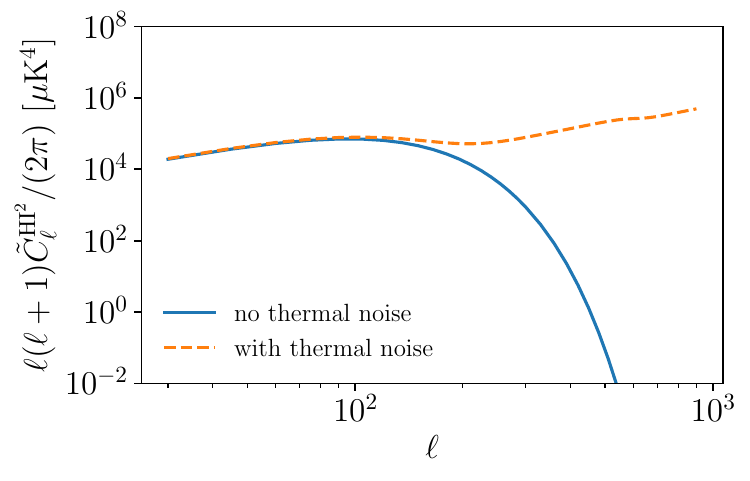}}
    \caption{Power spectrum of the squared fields, in which the biases $b_\mathrm{e}$ and $b_\mathrm{HI}$ are set to be unity. {\it Left}: Power spectrum of the squared filtered kSZ field (Eq.~(\ref{eq:Cell_kSZ2})) in which both ACT (blue solid line) and SO (orange solid line) are shown. {\it Right}: Power spectrum of the squared HI IM field with (orange dashed line) and without (blue solid line) thermal noise from instruments (Eq.~(\ref{eq: Cl HI PS field})).}
    \label{fig:auto-correlation}
\end{figure*}

\begin{figure*}[t]
    \centerline{\includegraphics[height=0.3\linewidth]{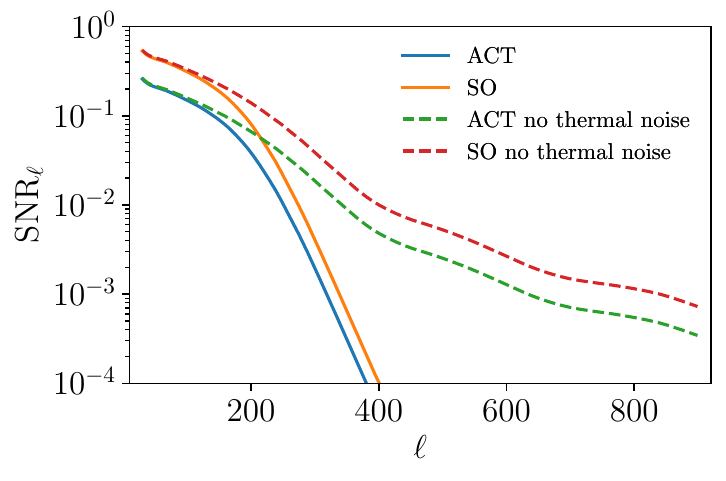}
    \includegraphics[height=0.3\linewidth]{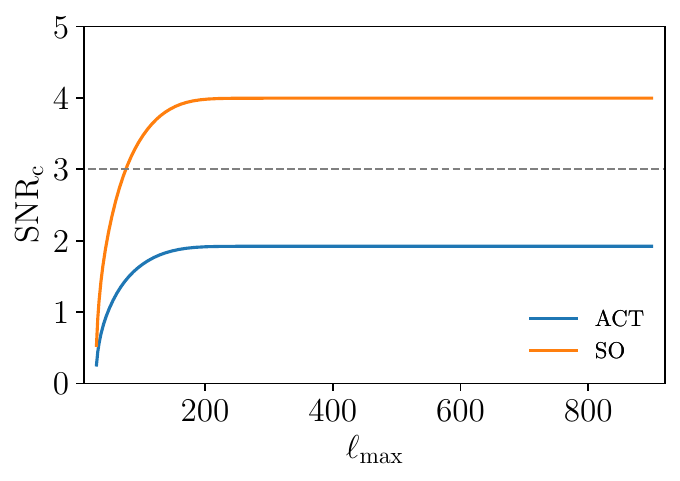}}
    \caption{Signal-to-noise ratio per-$\ell$ mode ($\mathrm{SNR}_\ell$, {\it left}) and cumulative SNR (SNR$_\mathrm{c}$, {\it right}) as functions of $\ell_{\mathrm{max}}$ for different CMB observations cross-correlating with SKA-MID, where in the right panel the horizontal dashed gray line denotes the benchmark ${\rm SNR}_{\rm c}=3$.} 
    \label{fig:SNR}
\end{figure*}

For $\tilde{C}_\ell^{\mathrm{HI}^2}$, the Gaussian approximation leads to a slightly different equation
\begin{align}
    \tilde{C}_\ell^{\mathrm{HI}^2} &= \int \md\chi\left(\frac{W_\mathrm{HI}}{\chi}\right)^2 \tilde{P}_{\mathrm{HI}^2}(k=\ell/\chi) \nonumber \\
    &= \int \md\chi \left(\frac{W_\mathrm{HI}}{\chi}\right)^2 2 \int\frac{\md^3 \vec{p}}{(2\pi)^3} \nonumber\\
    &\quad\quad\times B^{2}_{\mathrm{HI}}(p) B^{2}_{\mathrm{HI}}(|\vec{\ell}/\chi - \vec{p}|) \tilde{P}_{\mathrm{HI}}(p) \tilde{P}_{\mathrm{HI}}(|\vec{\ell}/\chi - \vec{p}|), \nonumber \\
    \label{eq: Cl HI PS field}
\end{align}
with
\begin{align}
    \tilde{P}_{\mathrm{HI}}(k) =  b_{\mathrm{HI}}^2 \left(1 + b_{\mathrm{HI}}^{-1}f\mu^2 \right)^2 P_{\mathrm{m}}(k) + B_{\mathrm{HI}}^{-2} P_{\mathrm{noise}}(k)~, 
\end{align}
where $P_{\mathrm{noise}}(k)$ is the 1D noise power spectrum for HI observation. Although both $\tilde{C}_\ell^{\mathrm{kSZ}^2}$ and $\tilde{C}_\ell^{\mathrm{HI}^2}$ contain convolutions of the power spectrum, they behave differently because the kSZ field is a projection field with no parallel mode information, but HI field is not. The noise mainly consists of the thermal noise from instruments, i.e. $P_{\mathrm{noise}}=P_\mathrm{T}$, whose power spectrum can be evaluated as~\cite{Bull:2014rha,Jiang:2023zex}
\begin{align}\label{eq: thermal noise power}
\begin{aligned}
    P_{\mathrm{T}} &=~ \sigma_{\mathrm{T}}^2  V_{\mathrm{voxel}} \\
    & =~ \frac{T_{\mathrm{sys}}^2}{\delta\nu t_{\mathrm{tot}}} \left(\frac{\lambda^2}{A_{\rm e}\theta_{\rm b}^2}\sqrt{\frac{A_{\rm s}}{\theta_{\rm b}^2}}~\right)^2 (\chi\theta_{\rm b})^2 \delta\chi~,
\end{aligned}
\end{align}
where the overall system temperature would be $T_{\mathrm{sys}}\simeq 22~\mathrm{K}$ (at $1200~\mathrm{MHz}$)~\cite{SKA:2018ckk}, $A_{\rm e}$ is the effective collecting area of a dish, $A_{\rm s}$ is the survey area, and $V_{\mathrm{voxel}}=(\chi\theta_{\mathrm{b}})^2\delta\chi$ is the comoving volume of each voxel. Here $\theta_\mathrm{b}$ is the FWHM of  the antenna and comoving distance resolution $\delta\chi$ at a given redshift is evaluated via Eq.~\eqref{eq:dchi_z} with frequency resolution $\delta \nu =0.2~\mathrm{MHz}$. 
The total observation time $t_{\mathrm{tot}} = 10^4\,{\rm hrs}$ for each dish leads to a relatively small enough noise power $P_{\mathrm{T}}\sim \mathcal{O}(1)~\mu\mathrm{K}^2$ that can be neglected comparing with the HI signal at large scale, suggesting a good approximation $\tilde{P}_{\mathrm{HI}}(k) \simeq P_{\mathrm{HI}}(k)$ for $\ell \leq 200$. The power spectra of the kSZ square field and 21-cm-squared field are show in the left and right panels of Fig.~\ref{fig:auto-correlation} respectively.

We finally show the signal-to-noise ratio (SNR) per-$\ell$ (${\rm SNR}_{\ell}$) and the total SNR as a function of $\ell_{\rm max}$ in the left and right panels of Fig.~\ref{fig:SNR}. We have set the sky fraction to be $f_{\mathrm{sky}}\simeq 0.2$ for the overlapping area between SKA-MID~\cite{SKA:2018ckk} and ACT~\cite{ACT:2025xdm}. As we can see, with $\ell_{\mathrm{max}}\simeq 200$ the cumulative SNR can arrive at a maximum value $\mathrm{SNR}_{\mathrm{max}}\simeq 1.92$. 
As a comparison, we show in Fig.~\ref{fig:SNR} the cumulative SNR for ACT and SO as functions of $\ell_\mathrm{max}$ respectively, the latter of which reaches a maximum SNR at $3.99$. 

In contrast to the cross-correlation signal itself, the SNR itself is insensitive to $b_\mathrm{HI}$ and $\Omega_\mathrm{HI}$ as long as the 21-cm detector noise is low enough. This can be understood by taking the noise-free limit of Eq.~\eqref{eq:SNR2}, where both the numerator and denominator are proportional to $b_\mathrm{HI}^2\Omega_\mathrm{HI}^2$ and thus cancel out. Nevertheless, SNR will be sensitive to $b_\mathrm{e}$ that comes from the kSZ effect, because the signal is filtered out from a noise-dominated background. In linear regime, the biases of both electron and HI are of order $\mathcal{O}(1)$~\cite{Villaescusa-Navarro:2018vsg,Jolicoeur:2020eup}, so the assumption of the unit biases is reasonable.

One of the most important factors that affect the SNR is the thermal noise of the HI observation. As is shown in the left panel of Fig.~\ref{fig:SNR}, the SNR per-$\ell$ begins to drop rapidly at the scale where thermal noise begins to dominate the autocorrelation of the HI square field. It should be noted that it is not the total observation time but the time per area that affects the noise power, as a proportion relationship $P_\mathrm{T} \propto A_\mathrm{s}/t_\mathrm{tot}$ is suggested from the definition Eq.~\eqref{eq: thermal noise power}. For simplicity let us assume the $10,000$ hours integration time mentioned is uniformly distributed into $20,000~\mathrm{deg}^2$. As the HI signal is going to cross-correlate with the kSZ observation with an overlap sky-coverage $8,000~\mathrm{deg}^2$, the effective integration time is actually $4,000$ hours.

At last, here we provide a rough estimation on the results with a shorter integration time, for example $1,000$ hours for $8,000~\mathrm{deg}^2$ sky-coverage, in which case the thermal noise would begin to dominate the HI square field autosignal at a larger scale $\ell\simeq 100$, leading to a suppression of SNR per mode starting at a smaller $\ell$. Then the approximate cumulative SNR would reach its maximum at an earlier stage. Let us truncate the cumulation at $\ell\simeq 100$, and then the results finally arrive at $\mathrm{SNR}_\mathrm{max}\simeq 3$ for SO and $\mathrm{SNR}_\mathrm{max}\simeq 1.5$ for ACT.

\section{Conclusions and discussions}\label{sec:con}

In this paper, we established the analytical formulas for the cross-correlation power spectrum between the kSZ square field and the projection of the HI square field under the Gaussian and flat-sky approximations. In terms of the validity of the approximation, for the low-redshifts HI intensity mapping field ($z\lesssim 3$) where reionizations of hydrogen and helium are fully completed, the Gaussian approximation works well for the perturbed HI intensity field that follows the matter density field. 
However, we caution that these formulas may be inaccurate at higher redshifts during reionization ($6\lesssim z \lesssim 15$), where the ionized regions lead to a strongly non-Gaussian signal while spin temperature fluctuations may lead to further complications, at least during early phases of the EoR~\cite{Mesinger2011}. 
In that scenario, the analytical formula we can achieve at most is Eq.~\eqref{eq: C_l vector}, in which multipoint correlation $\mathcal{T}$ must be evaluated in a fully numerical way. For the flat-sky approximation, it is reasonable as long as we consider high-$\ell$ modes of correlation, in which the curvature of the sky can be safely neglected. To calculate the complete, full-sky power spectrum, one needs to work with the spherical-harmonics decomposition instead of 2D plane-wave Fourier transformation, which is left to further study. 

To demonstrate the results, we numerically calculated the cross-correlation power spectrum in the range of $30\leq \ell \leq 900$, between the redshift $0.3\leq z \leq 1$ which is the sensitive range of the SKA-MID IM project. We forecast the SNR for SKA-MID to reach a maximum value $\mathrm{SNR}_{\mathrm{max}}\simeq 1.92$ for ACT and $\mathrm{SNR}_{\mathrm{max}}\simeq 3.99$ for SO. These results come from the assumption of constant biases for the electron and HI density contrasts, which is reasonable because the main contributions in Eq.~\eqref{eq: C_l simplified} come from the peak of the power spectrum, where the nonlinear effects are safely neglected. Future investigations can be done to calculate the scale-dependent biases derived from the halo model.

One of the advantages of taking the square field cross-correlation is that the convolution arising from the square field helps to leverage the $k^{\pll}$ modes in $2$D map cross-correlations, catching more information about the large-scale structure of the universe than simple projection fields. The $k^{\pll}$ modes will become more important when the evolution of the structure is not negligible, for example, during the EoR period. Future studies can be carried out by evaluating the $\mathrm{kSZ}^2\times\mathrm{HI}^2$ cross-correlation during EoR for HI signal detection and exploring the physics of the first stars and galaxies at high redshifts (e.g. Ref.~\cite{Zhou:2025fgv}).

Here we briefly discuss possible methods to improve the SNR from the perspectives of observations and instruments. The primary factors affecting SNR are noise levels and beams, which would affect the value of Wiener filter  response and also lead to the noise power dominating over the signal on small scales. The most effective way to boost the SNR should be observing the 21-cm signal with a finer beam, allowing more small-scale information surviving from beam damping. As being discussed in Appendix~\ref{app: power no beam effect}, a finer beam significantly enhances the SNR by allowing $C^{(7)}_{\ell}$ to dominate the cross-correlation at large $\ell$s, which corresponds to the contribution from velocity autocorrelation $P(k)$ at the $k\to 0$ limit.

As a final remark, the cross-correlations between HI intensity mapping and other tracers of large-scale structure can serve as a validation of methodology for high-redshift 21-cm detection, which faces severe challenges from foreground contamination, especially when coupled with instrumental effects (e.g. the mutual coupling effect~\cite{Charles2024,Rath2025} and foreground leakage~\cite{Orosz2019,Gogo2022}). 
The kSZ signal during the EoR is also one of the few accessible signals across large regions of the sky. At low redshifts, we carried out a detailed study on the cross-correlation of the kSZ square field and HI square field, which provides a presumably easier test of the statistics. A validation of this method at low redshifts can provide further motivation to pursue similar procedure back into the EoR. In addition, at lower redshifts, the expected kSZ signals are better understood, and there are already detected cross-correlations with galaxy surveys~\cite{Hand2012,Li2018,Planck2016-unbound} and velocity reconstruction fields~\cite{Planck2016-unbound,CHM2015}. Our method opened a new window by novelly introducing their cross-correlation with a 21-cm signal, providing a testing ground and cross-check for other detection statistics.

\acknowledgments
The authors acknowledge the useful discussions with Meng Zhou, Steven Murray and the MeerKLASS collaboration team. Z.-Y. Yuwen and Y.-E. Jiang are supported by the Program of China Scholarship Council Grant No. 202404910329 and No. 202404910398. Y.-Z. Ma acknowledges the support from South Africa's National Research Foundation under Grant No. 150580, No. CHN22111069370, No. ERC250324306141. P. L. Plante is supported by NSF Grant No. 2206602 and the Simons Foundation award No. 00007127. All computations were carried out using the computational cluster resources at the Centre for High-Performance Computing, Cape Town, South Africa.

\appendix

\section{Projected square field v.s. Square of projected field}\label{app: PS v.s. SP}

In this appendix, we provide detailed comparison between the cross-correlations of the projected square field and the square of projected field with other tracers, which provide insights to the methodology (square of the projected field) further. While this work employs intensity mapping as an example, the presented methodology maintains general applicability to any tomographic survey. Consider the following two slightly different procedures for dealing with the HI signal: 
\begin{itemize}
    \item $\delta T_{\mathrm{HI}}^{\mathrm{SP}}(\hat{n})$, where ``SP'' stands for the square of projected field, i.e. first projecting the 3D field into a 2D field, then square the field;    
    \item $\delta T_{\mathrm{HI}}^{\mathrm{PS}}(\hat{n})$, where ``PS'' stands for the projection of the square field, i.e. first squaring the entire 3D volume, and then projecting it into a 2D field.
\end{itemize}
Let us investigate their 2D Fourier modes respectively, and the cross-correlations with another 2D field [e.g. the galaxy field $\delta_{\mathrm{g}}(\hat{n})$], to compare the extractable information.

For the SP field, one can first get the projection of the HI signal along the line-of-sight given by Eq.~\eqref{eq:projection2D} and then square it, from which the SP field reads
\begin{align}
\begin{aligned}
    \delta T_{\mathrm{HI}}^{\mathrm{SP}}(\hat{n}) &= \left(\int \md\chi ~ W_{\mathrm{HI}}(\chi) \bar{T}_{\mathrm{b}}(\chi) \delta_{\mathrm{HI}}(\chi\hat{n},\chi)\right)^2 \\
    &= 
    \left(\int \md\chi ~ K_{\mathrm{HI}}(\chi) \int \frac{\md^3\vec{k}}{(2\pi)^3}\me^{i\vec{k}\cdot\vec{x}}\delta_{\mathrm{HI}}(\vec{k}) 
    \right)^2,
\end{aligned}
\end{align}
with abbreviation $K_{\mathrm{HI}}(\chi)\equiv W_{\mathrm{HI}}(\chi) \bar{T}_{\mathrm{b}}(\chi)$. Here we omit the beam function for simplicity. Performing a 2D Fourier transformation over the direction $\hat{n}$ results in
\begin{widetext}
\begin{align}
    \delta T_{\mathrm{HI}}^{\mathrm{SP}}(\vec{\ell}) &= \int\md^2\vec{\theta} ~ \me^{-i\vec{\ell}\cdot \vec{\theta}} \delta T_{\mathrm{HI}}^{\mathrm{SP}}(\hat{n}) 
    \nonumber\\
    &= \int\md\chi_1\md\chi_2 ~ 
    K_{\mathrm{HI}}(\chi_1)K_{\mathrm{HI}}(\chi_2) 
    \int \frac{\md^3\vec{k}_1 \md^3 \vec{k}_2}{(2\pi)^6} \delta_{\mathrm{HI}}(\vec{k}_1) \delta_{\mathrm{HI}}(\vec{k}_2)
    \int\md^2\vec{\theta}~\me^{i((\vec{k}_1+\vec{k}_2) \cdot\vec{x}-\vec{\ell}\cdot \vec{\theta} )} 
    \nonumber\\
    & = \int\md\chi_1\md\chi_2 ~ K_{\mathrm{HI}}(\chi_1)K_{\mathrm{HI}}(\chi_2) \int\frac{\md k_1^{\pll} \md k_2^{\pll}}{(2\pi)^2}~\me^{i(k_1^{\pll}\chi_1 + k_2^{\pll}\chi_2)} 
    \int \frac{\md^2 \vec{k}_1^\perp \md^2 \vec{k}_2^\perp}{(2\pi)^4}(2\pi)^2 \delta_\mathrm{D}^{(2)}(\vec{k}_1^\perp\chi_1 + \vec{k}_2^\perp\chi_2 - \vec{\ell}) \delta_{\mathrm{HI}}(\vec{k}_1) \delta_{\mathrm{HI}}(\vec{k}_2) 
    \nonumber\\
    & = \int\md\chi_1\md\chi_2 ~ \frac{K_{\mathrm{HI}}(\chi_1) K_{\mathrm{HI}}(\chi_2)}{\chi_2^2} \int\frac{\md k_1^{\pll} \md k_2^{\pll}}{(2\pi)^2}~\me^{i(k_1^{\pll}\chi_1 + k_2^{\pll}\chi_2)} 
    \int \frac{\md^2 \vec{k}_1^\perp}{(2\pi)^2} \delta_{\mathrm{HI}}\left(\vec{k}_1^\perp, k_1^{\pll}\right) \delta_{\mathrm{HI}}\left(\frac{\vec{\ell}}{\chi_2} - \frac{\chi_1}{\chi_2}\vec{k}_1^\perp, k_2^{\pll}\right) ~,
\end{align}
where we have used the same fact given in Eq.~\eqref{eq: delta_D^2 of kperp} to integrate over $\vec{\theta}$.

By definition, the PS field is given by the projection of the square of 3D HI field $\delta_{\mathrm{HI}^2}(\vec{x}) \equiv \delta_{\mathrm{HI}}(\vec{x})^2$, along line-of-sight
\begin{align}
    \delta T_{\mathrm{HI}}^{\mathrm{PS}}(\hat{n}) &= \int \md\chi ~ W_{\mathrm{HI}}(\chi) \bar{T}_{\mathrm{b}}(\chi)^2 \delta_{\mathrm{HI}^2}(\chi\hat{n},\chi) = \int \md\chi ~ K_{\mathrm{HI}^2}(\chi)
    \int \frac{\md^3\vec{k}}{(2\pi)^3} \me^{i\vec{k}\cdot\vec{x}} ~\delta_{\mathrm{HI}^2}(\vec{k}) \nonumber \\
    & = \int \md\chi ~ K_{\mathrm{HI}^2}(\chi) \int \frac{\md^3\vec{k}}{(2\pi)^3} \me^{i\vec{k}\cdot\vec{x}} \int\frac{\md^3\vec{k}'}{(2\pi)^3} ~\delta_{\mathrm{HI}}(\vec{k} - \vec{k}') \delta_{\mathrm{HI}}(\vec{k}'),
\end{align}
with $K_{\mathrm{HI}^2}(\chi) \equiv W_{\mathrm{HI}}(\chi) \bar{T}_{\mathrm{b}}(\chi)^2$, and the Fourier modes $\delta_{\mathrm{HI}^2}(\vec{k})$ expressed in a convolution of two copies of $\delta_{\mathrm{HI}}(\vec{k})$. Then the 2D Fourier transformation leads to
\begin{align}
    \delta T_{\mathrm{HI}}^{\mathrm{PS}}(\vec{\ell}) &= \int\md^2\vec{\theta} ~ \me^{-i\vec{\ell}\cdot \vec{\theta}} \delta T_{\mathrm{HI}}^{\mathrm{PS}}(\hat{n}) 
    \nonumber\\
    &= \int \md\chi ~ K_{\mathrm{HI}^2}(\chi) \int \frac{\md^3\vec{k}}{(2\pi)^3} 
    \int\frac{\md^3\vec{k}'}{(2\pi)^3} ~\delta_{\mathrm{HI}}(\vec{k} - \vec{k}') \delta_{\mathrm{HI}}(\vec{k}') \int\md^2\vec{\theta}~\me^{i\vec{k}\cdot\vec{x} - i\vec{\ell}\cdot \vec{\theta}} 
    \nonumber\\
    &= \int \md\chi ~ K_{\mathrm{HI}^2}(\chi) \int \frac{\md k^{\pll}}{2\pi} \me^{ik^{\pll}\chi} \int\frac{\md^3\vec{k}'}{(2\pi)^3} ~\delta_{\mathrm{HI}}(\vec{k} - \vec{k}') \delta_{\mathrm{HI}}(\vec{k}') \int \frac{\md^2\vec{k}^\perp}{(2\pi)^2} (2\pi)^2 \delta_\mathrm{D}^{(2)}(\vec{k}^\perp \chi - \vec{\ell})
    \nonumber\\
    &= \int \md\chi ~ \frac{K_{\mathrm{HI}^2}(\chi)}{\chi^2} \int \frac{\md k^{\pll}}{2\pi} \me^{ik^{\pll}\chi} \int\frac{\md^3\vec{k}'}{(2\pi)^3} ~\delta_{\mathrm{HI}}(\vec{k}^{\prime \perp}, k^{\prime \pll}) \delta_{\mathrm{HI}} \left( \frac{\vec{\ell}}{\chi}- \vec{k}^{\prime \perp}, k^{\pll} - k^{\prime \pll} \right). 
\end{align}

The projected galaxy field and its 2D Fourier transformation can be derived from Eq.~\eqref{eq:projection2D} and Eq.~\eqref{eq: Xl} respectively, by replacing $X$ with $g$
\begin{align}
    \delta_{\mathrm{g}}(\hat{n}) &= \int \md\chi ~ W_{\mathrm{g}}(\chi) \delta_{\mathrm{g}}(\chi\hat{n},\chi) = \int \md\chi ~ W_{\mathrm{g}}(\chi) \int \frac{\md^3\vec{k}}{(2\pi)^3}\me^{i\vec{k}\cdot\vec{x}}\delta_{\mathrm{g}}(\vec{k}) ~, \\
    \delta_{\mathrm{g}}(\vec{\ell}) &= \int \md\chi ~ \frac{W_{\mathrm{g}}(\chi)}{\chi^2} \int \frac{\md k^{\pll}}{2\pi} \me^{ik^{\pll}\chi} \delta_{\mathrm{g}}(\vec{\ell}/\chi,k^{\pll}).
\end{align}
Then the cross-correlation between $\delta T_{\mathrm{HI}}^{\mathrm{SP}}$ and $\delta_{\mathrm{g}}$ is given by
\begin{align} \label{eq: C_l SP cross g}
    C_\ell^{\mathrm{SP}\times {\mathrm{g}}} &= \int \frac{\md^2 \vec{\ell}'}{(2\pi)^2} \left\langle \delta T_\mathrm{\mathrm{SP}}(\vec{\ell}) \delta_{\mathrm{g}}(\vec{\ell}') \right\rangle 
    \nonumber \\
    &= \int \frac{\md^2 \vec{\ell}'}{(2\pi)^2} \int\md\chi_1\md\chi_2\md\chi_3 ~  K_{\mathrm{HI}}(\chi_1) \frac{K_{\mathrm{HI}}(\chi_2)}{\chi_2^2} \frac{W_{\mathrm{g}}(\chi_3)}{\chi_3^2} 
    \int\frac{\md k_1^{\pll} \md k_2^{\pll} \md k_3^{\pll}}{(2\pi)^3}~\me^{i(k_1^{\pll}\chi_1 + k_2^{\pll}\chi_2 + k_3^{\pll}\chi_3)} 
    \nonumber\\
    &\quad\quad\quad\times \int \frac{\md^2 \vec{k}_1^\perp}{(2\pi)^2} \left\langle\delta_{\mathrm{HI}}\left(\vec{k}_1^\perp, k_1^{\pll}\right) \delta_{\mathrm{HI}}\left(\frac{\vec{\ell}}{\chi_2} - \frac{\chi_1}{\chi_2}\vec{k}_1^\perp, k_2^{\pll}\right) \delta_{\mathrm{g}} \left(\frac{\vec{\ell}'}{\chi_3},k_3^{\pll}\right) \right\rangle~.
\end{align}
In comparison, replacing the SP field with the PS field suggests a different cross-correlation
\begin{align} \label{eq: C_l PS cross g}
    C_\ell^{\mathrm{PS}\times {\mathrm{g}}} &= \int \frac{\md^2 \vec{\ell}'}{(2\pi)^2} \left\langle \delta T_\mathrm{\mathrm{PS}}(\vec{\ell}) \delta_{\mathrm{g}}(\vec{\ell}') \right\rangle 
    \nonumber \\
    &= \int \frac{\md^2 \vec{\ell}'}{(2\pi)^2} \int\md\chi_1\md\chi_2 ~  \frac{K_{\mathrm{HI}^2}(\chi_1)}{\chi_1^2} \frac{W_{\mathrm{g}}(\chi_2)}{\chi_2^2} 
    \int\frac{\md k_1^{\pll} \md k_2^{\pll}}{(2\pi)^2}~\me^{i(k_1^{\pll}\chi_1 + k_2^{\pll}\chi_2)} 
    \nonumber\\
    &\quad\quad\quad\times \int\frac{\md^3\vec{k}'}{(2\pi)^3} 
    \left\langle \delta_{\mathrm{HI}}(\vec{k}^{\prime \perp}, k^{\prime \ell}) \delta_{\mathrm{HI}} \left( \frac{\vec{\ell}}{\chi_1}- \vec{k}^{\prime \perp}, k_1^{\pll} - k^{\prime \ell} \right) \delta_{\mathrm{g}} \left(\frac{\vec{\ell}'}{\chi_2},k_2^{\pll}\right) \right\rangle~.
\end{align}
Each of these two cross-correlations contains a bispectrum term $\mathcal{B}^{\mathrm{HI}^2 \mathrm{g}}(\vec{k}_1,\vec{k}_2,\vec{k}_3)$ with a closed triangle constraint $\vec{k}_1+\vec{k}_2+\vec{k}_3=0$
\begin{align}
    \left\langle \delta_{\mathrm{HI}}(\vec{k}_1) \delta_{\mathrm{HI}}(\vec{k}_2) \delta_{\mathrm{g}}(\vec{k}_3) \right\rangle = (2\pi)^3 \delta_D^{(3)}(\vec{k}_1 + \vec{k}_2 + \vec{k}_3) \mathcal{B}^{\mathrm{HI}^2 \mathrm{g}}(\vec{k}_1,\vec{k}_2,\vec{k}_3)~.
\end{align}
We now apply Limber approximation, which assumes a slow-varying parallel modes dependence in the bispectrum for a small-angular scale satisfying $k^{\pll}\ll \ell/\chi$ (as long as the mode is not enrolled in the convolution). This allows us to neglect the $k^{\pll}$ in $\mathcal{B}^{\mathrm{HI}^2 \mathrm{g}}$ and simplify Eqs.~\eqref{eq: C_l SP cross g} and~\eqref{eq: C_l PS cross g} to the following expressions
\begin{eqnarray}
    C_\ell^{\mathrm{SP}\times {\mathrm{g}}} &=& \int \md\chi~ \bar{T}_{\mathrm{b}}(\chi)^2\frac{W_{\mathrm{HI}}(\chi)^2 W_{\mathrm{g}}(\chi)}{\chi^2} \int \frac{\md^2 \vec{k}^\perp}{(2\pi)^2} ~\mathcal{B}^{\mathrm{HI}^2 \mathrm{g}} \left(\vec{k}^\perp,\frac{\vec{\ell}}{\chi} - \vec{k}^\perp, -\frac{\vec{\ell}}{\chi}\right)~,\label{eq:SP-g1} \\
    C_\ell^{\mathrm{PS}\times {\mathrm{g}}} &=& \int \md\chi~ \bar{T}_{\mathrm{b}}(\chi)^2\frac{W_{\mathrm{HI}}(\chi) W_{\mathrm{g}}(\chi)}{\chi^2} \int \frac{\md^3 \vec{k}}{(2\pi)^3} ~ \mathcal{B}^{\mathrm{HI}^2 \mathrm{g}} \left(\vec{k},\frac{\vec{\ell}}{\chi} - \vec{k}, -\frac{\vec{\ell}}{\chi}\right)~, \label{eq:PS-g1}
\end{eqnarray}
\end{widetext}
where $\cos\theta = \hat{k}\cdot\hat{\ell}$. 
The last line is derived by noticing that the bispectrum should be a function of $|\vec{k}|$, $\ell/\chi$ and the separation angle $\cos\theta$. A similar result remains correct if one replaces $\delta_{\mathrm{g}}$ with another projected field such as kSZ, as is given in Eq.~\eqref{eq: Cli}.

One can clearly see the difference between Eqs.~(\ref{eq:SP-g1}) and (\ref{eq:PS-g1}), which is the momentum integral of the bispectrum. The PS method results in a 3D integral which fully takes the advantage of parallel momentum mode, whereas SP method loses this $k^{\parallel}$ information and only utilizes the $k^{\perp}$ modes. Therefore, the $C_\ell^{\mathrm{PS}\times {\mathrm{g}}}$ has significantly higher signal than $C_\ell^{\mathrm{SP}\times {\mathrm{g}}}$.

\section{Redshift-space distortion effect of the HI intensity mapping}
\label{app: RSD}

\begin{figure*}[t]
\centerline{\includegraphics[height=0.32\linewidth]{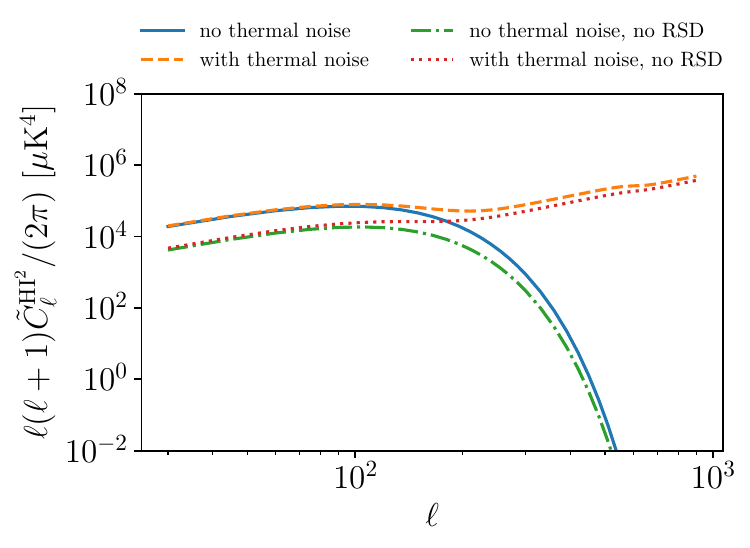}
\includegraphics[height=0.325\linewidth]{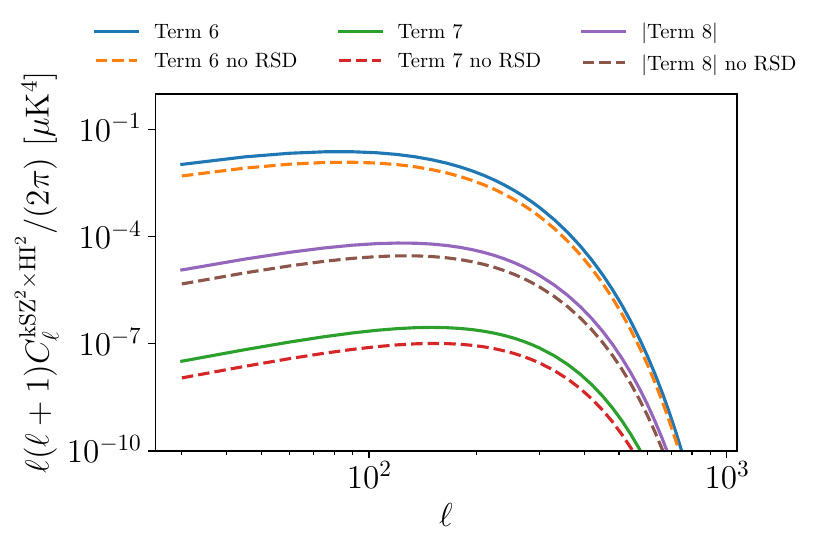}}
\caption{Correlation functions with and without RSD effect. {\it Left}--Power spectrum of the HI square field; {\it Right}--Each individual term contribution in the cross-power spectrum between ${\rm kSZ}^{2}$ and ${\rm HI}^{2}$ fields. The absolute value of total cross-correlation $C_\ell^\mathrm{tot}$ coincides with $C_\ell^{(6)}$ because of the hierarchies among contributions from different terms.}
\label{fig:RSD}
\end{figure*}

In this appendix, we discuss how the redshift-space distortion (RSD) correction on HI intensity mapping is applied to the density contrasts (and further to the correlations functions). In a realistic HI intensity mapping observation, the observable is a radiative quantity -- the brightness $I_\nu$, which is defined by the received energy $\md E$ per unit time $\md t$ per detector area $\md A$ per solid angle $\md \Omega$ within the frequency range $(\nu, \nu+\md \nu)$
\begin{align}
    I_\nu = \frac{\md E}{\md t \md A \md\Omega \md \nu} \equiv \frac{F_\nu}{\md \Omega \md\nu}~,
\end{align}
where $F_\nu=I_\nu \md \Omega\md\nu$ is the flux onto a detector at frequency $\nu$ as the physical observable quantity~\cite{Shaw:2008aa}. Since the redshift of the source is determined by the shift from its original frequency $\nu_0$, i.e. $z=\nu_0/\nu - 1$, the \textit{redshift-space} comoving distance in the range $(\nu,\nu+\md\nu)$ is given by
\begin{eqnarray}
    \md\chi_z = \frac{c}{H}\md z = -\frac{c}{H}\frac{\nu_0\md\nu}{\nu^2} ~. \label{eq:dchi_z}
\end{eqnarray}
Equation~(\ref{eq:dchi_z}) may not be equal to the actual comoving distance $\chi$ in a perturbed universe because of the peculiar velocity of the source. In the presence of peculiar velocities, the comoving distance is given by
\begin{align}\label{eq: chis and chi}
    \chi = \chi_z - \frac{\hat{n}\cdot\vec{v}}{\mathcal{H}}\bigg|_\chi~,
\end{align}
where $\mathcal{H} = aH$ is the comoving Hubble parameter. We have neglected the relatively small local evolution of the background, lensing and general-relativistic effects, for which more details can be found in Ref.~\cite{Hall:2012wd}. The energy conservation then reads
\begin{align}\label{eq: brightness dOmega dchi}
    I_{s,\nu}(\chi_s)\md\Omega\md\chi_s =     I_\nu(\chi)\md\Omega\md\chi~,
\end{align}
where the subscript $s$ denotes quantities in redshift space. In the Rayleigh-Jeans regime, the brightness temperature is proportional to the $I_{\nu}$ function, i.e. $T_\mathrm{b} \simeq I_\nu c^2 / (2k_\mathrm{B}\nu^2)$, and one can then rewrite Eq.~\eqref{eq: brightness dOmega dchi} as
\begin{align}
    \bar{T}_b(1+\delta_{s,\mathrm{HI}}) \md\Omega\md\chi_s =  \bar{T}_b(1+\delta_{\mathrm{HI}}) \md\Omega\md\chi~.
\end{align}
Here we have used the fact that the perturbation of brightness temperature $\delta_{T_\mathrm{b}}$ equals the density contrast of the HI $\delta_{\mathrm{HI}}$. Further using the relation \eqref{eq: chis and chi} results in
\begin{align}
    \delta_{s,\mathrm{HI}} = \frac{\delta_{\mathrm{HI}} - \hat{n}\cdot\partial_\chi \vec{v}/\mathcal{H}}{1 + \hat{n}\cdot\partial_\chi \vec{v}/\mathcal{H}} ~.
\end{align}
In Fourier space, the relation between perturbations in redshift space and real space to the leading order reads
\begin{align}
    \delta_{s,\mathrm{HI}}(\vec{k}) = \delta_{\mathrm{HI}}(\vec{k}) - \frac{ik^{\pll}}{\mathcal{H}} \hat{n}\cdot \vec{v}(\vec{k})~.
\end{align}
Substitute HI perturbation with the matter perturbation together with a bias $b_{\mathrm{HI}}$, and replace velocity with Eq.~\eqref{eq: velocity of k}
\begin{align}
    \delta_{s,\mathrm{HI}}(\vec{k}) = \left(b_{\mathrm{HI}} + f\mu^2 \right)\delta_\mathrm{m}(\vec{k}) ~,
\end{align}
with abbreviation $\mu = k^{\pll}/k = \hat{n}\cdot\hat{k}$.

While evaluating the power spectra, the presence of the RSD effect means the replacement: $\delta_{\mathrm{HI}}\to \delta_{s,\mathrm{HI}}$. Then the HI field auto correlation power spectrum now becomes
\begin{align}
    P_{s,\mathrm{HI}}(\vec{k}) = b_{\mathrm{HI}}^2 \left(1 + b_{\mathrm{HI}}^{-1}f\mu^2 \right)^2 P_{\mathrm{HI}}(k),
\end{align}
which is nothing but the classical Kaiser effect~\cite{Kaiser:1984}. The same replacement can be applied to other correlation functions. The inclusion of RSD effect in auto- and cross-correlation power spectra are shown in Fig.~\ref{fig:RSD}. 
Similar to HI autocorrelation case, the amplitude of the power spectrum is increased about $1.5$ times with the inclusion of the RSD effect, while the shape of it remains not affected.

\section{Cross-correlation power spectrum in the fine-beam limit $\theta^{\rm HI}_{\rm FWHM}\rightarrow 0$}
\label{app: power no beam effect}

\begin{figure}
    \centering
    \includegraphics[width=0.9\linewidth]{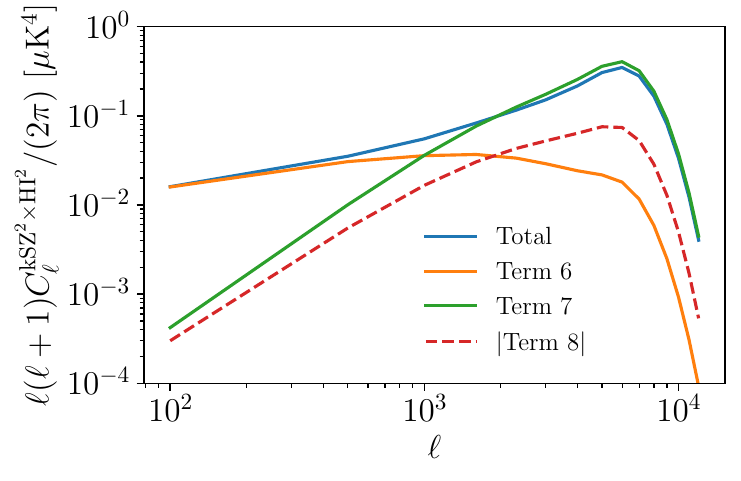}
    \caption{Cross-correlation angular power spectrum for HI IM in the redshift range $0.3<z<0.6$ in the fine-beam limit ($\theta^{\rm HI}_{\rm FWHM}\rightarrow 0$), with biases $b_\mathrm{e}=b_\mathrm{HI}=1$.}
    \label{fig: fine-beam limit}
\end{figure}

It is obvious that the cross-correlation power spectrum [Eq.~\eqref{eq: C_l simplified}] is suppressed by both the kSZ filter and the HI beam. The former is hard to modify on large scales because it comes from the ratio of power between the kSZ effect and primary CMB, which is order of $10^{-8}$ in general. However, the latter can be improved by enlarging the observation baseline, which can be achieved by using a much larger dish such as FAST telescope with effective diameter $D=300\mathrm{m}$~\cite{Nan:2011um}, or the interferometry experiments with longer baselines~\cite{SKA:2018ckk}. Here we briefly discuss the fine-beam limit of the HI ($\theta^{\rm HI}_{\rm FWHM}\rightarrow 0$, i.e. $B_\mathrm{HI}\to 1$) to see the capacity of signal enhancement.

We first recall that $Y^{(7)}$ given by Eq.~\eqref{eq: Yi results} contains the $P_{v,v}(k)$ factor, which has a ``resonant divergence'' as $k\to 0$, corresponding to the condition $\vec{l}_1+\vec{p}\chi\to 0$. For a rough estimation, one can neglect the change of the other two $P_\mathrm{e,HI}$ factors in the vicinity of the pole $\vec{l}_1+\vec{p}=0$. However, such a contribution from the pole will be highly suppressed at large $p^\perp = \ell_1/\chi$ with $p^{\pll}=0$, since it has been killed by the HI beam. 
The fine-beam limit allows high $p_{\perp}$ to survive, making a contribution to an effective factor proportional to the root-mean-square of velocity perturbation. Then the integral over $\vec{p}$ can be written as
\begin{align}
    &\int\frac{\md^3\vec{p}}{(2\pi)^3} Y^{(7)} \nonumber\\
    &\simeq P_{\mathrm{e},\mathrm{HI}}\left(\left|
    \frac{\vec{\ell} -\vec{\ell}_1}{\chi}
    \right|\right) P_{\mathrm{e},\mathrm{HI}} \left(\frac{\ell_1}{\chi}\right) \int\frac{\md^3\vec{k}}{(2\pi)^3}\mu^2
    P_{v,v}\left(k\right)  \nonumber\\
    &\equiv \frac{v_\mathrm{rms}^2}{3} P_{\mathrm{e},\mathrm{HI}}\left(\left|
    \frac{\vec{\ell} -\vec{\ell}_1}{\chi}
    \right|\right) P_{\mathrm{e},\mathrm{HI}} \left(\frac{\ell_1}{\chi}\right) ~.
\end{align}
Here we show the numerical results of the cross-correlation in Fig.~\ref{fig: fine-beam limit} within the redshift window $0.3<z<0.6$. 
From Fig.~\ref{fig: fine-beam limit}, we can see that although $C^{(6)}_{\ell}$ still dominate the contribution on large scales, $C^{(7)}_{\ell}$ gradually takes over as $\ell$ grows and eventually reaches a peak at $\ell\simeq 6000$, and then decays at smaller scales because of the kSZ filter. In this numerical calculation we did not include foreground filter to reach the limit $k^{\pll}\to 0$. By comparing Fig.~\ref{fig: fine-beam limit} with Fig.~\ref{fig:cross-correlation}, one can see that Term 7 and the total power spectrum can be enhanced to $\mathcal{O}(5\times 10^{-1})$ if $\ell\simeq 6000$ can be measured. However, this indicates that the dish size or interferometer baseline has to be at least $\sim 605$ m\footnote{We calculate the dish size by equating
\begin{eqnarray}
    \theta=\frac{\lambda}{D}=\frac{\pi}{\ell} \Rightarrow D=\frac{\lambda \ell}{\pi}=\left(\frac{c}{\nu} \right)\frac{\ell}{\pi}\simeq 605\,{\rm m}.
\end{eqnarray}
}, which excludes the possibility of single-dish measurement, but resorts to the interferometer measurements.

\bibliographystyle{apsrev}
\bibliography{ref}

\end{document}